\documentclass[useAMS,usenatbib,usegraphicx]{mn2e}
\usepackage{amsmath,amsfonts,amssymb}

\newcommand{\lya}{Ly$\alpha$}

\newcommand{\1}{$^{-1}$}
\newcommand{\2}{$^{-2}$}
\newcommand{\3}{$^{-3}$}
\newcommand{\hm}{$h^{-1}$}
\newcommand{\kms}{km s$^{-1}$}
\newcommand{\msun}{M$_{\sun}$}
\newcommand{\phot}{photon s$^{-1}$ cm$^{-2}$ sr$^{-1}$}
\newcommand{\K}{{\rm K}}

\newcommand{\cm}{{\rm cm}}

\newcommand{\aap}{A\&A}
\newcommand{\apj}{ApJ}
\newcommand{\apjl}{ApJL}
\newcommand{\apjs}{ApJS}
\newcommand{\mnras}{MNRAS}
\newcommand{\pasj}{PASJ}
\newcommand{\pasp}{PASP}
\newcommand{\nat}{Nature}
\newcommand{\araa}{ARA\&A}
\newcommand{\ssr}{SSR}
\newcommand{\apss}{Ap\&SS}
\newcommand{\spie}{SPIE}

\newcommand{\ion}[2]{#1\,{\sc{#2}}}
\newcommand{\hi}{\ion{H}{I}}
\newcommand{\ciii}{\ion{C}{III}}
\newcommand{\civ}{\ion{C}{IV}}
\newcommand{\siiv}{\ion{Si}{IV}}
\newcommand{\nv}{\ion{N}{V}}
\newcommand{\ovi}{\ion{O}{VI}}

\newcommand{\oviii}{\ion{O}{VIII}}
\newcommand{\neviii}{\ion{Ne}{VIII}}

\newcommand{\owls}{OWLS}
\newcommand{\cloudy}{{\small CLOUDY}}
\newcommand{\xray}{X-ray}
\newcommand{\hstcos}{{\it COS}}
\newcommand{\stis}{{\it STIS}}
\newcommand{\fuse}{{\it FUSE}}
\newcommand{\atlast}{{\it ATLAST}}
\newcommand{\fireball}{{\it FIREBALL}}

\newcommand{\default}{\emph{REF}}                % DEFAULT
\newcommand{\nosn}{\emph{NOSN}}                  % NOFB
\newcommand{\zcool}{\emph{NOZCOOL}}              % ZCOOL0
\newcommand{\wmom}{\emph{WVCIRC}}                % WMOM
\newcommand{\agn}{\emph{AGN}}
\newcommand{\mill}{\emph{MILL}}
\newcommand{\wdens}{\emph{WDENS}}
\newcommand{\wml}{\emph{WML4}}
\newcommand{\dblimf}{\emph{DBLIMF}}

\title[UV emission from the WHIM]{Metal-line emission from the warm-hot intergalactic medium: II. Ultraviolet}
\author[Bertone et al.]{Serena Bertone$^{1}$\thanks{E-mail: serena@scipp.ucsc.edu},
Joop Schaye$^{2}$,
C. M. Booth$^{2}$,
Claudio Dalla Vecchia$^{2,3}$,
\newauthor
Tom Theuns$^{4,5}$ and 
Robert P.~C. Wiersma$^{2,6}$
\\
$^{1}$Santa Cruz Institute for Particle Physics, University of California, 1156 High Street, Santa Cruz CA 95064, USA \\
$^{2}$Leiden Observatory, Leiden University, P.O. Box 9513, 2300 RA Leiden, The Netherlands \\
$^{3}$Max Planck Institute for Extraterrestrial Physics, Giessenbachstrasse 1, D-85748 Garching bei M\"unchen, Germany \\
$^{4}$Institute for Computational cosmology, Department of Physics, University of Durham, South Road, Durham, DH1 3LE \\
$^{5}$Universiteit Antwerpen, Campus Groenenborger, Groenenborgerlaan
171, B-2020 Antwerpen, Belgium \\
$^{6}$Max Planck Institut f\"ur Astrophysik, Karl Schwarzschild Str. 1, Postfach 1317, D-85741, Garching bei M\"unchen, Germany
}

\begin{document}
\date{Submitted to MNRAS}
\pagerange{\pageref{firstpage}--\pageref{lastpage}} \pubyear{2009}
\maketitle
\label{firstpage}

\begin{abstract}
Approximately half the baryons in the local Universe are thought to reside in the warm-hot intergalactic medium (WHIM), i.e.\ diffuse gas with temperatures in the range $10^5\,$K $<T<10^7\,$K. Emission lines from metals in the UV band are excellent tracers of the cooler fraction of this gas, with $T\la 10^6\,$K. We present predictions for the surface brightness of a sample of UV lines that could potentially be observed by the next generation of UV telescopes at $z<1$. We use a subset of simulations from the \owls\ project to create emission maps and to investigate the effect of varying the physical prescriptions for star formation, supernova and AGN feedback, chemodynamics and radiative cooling. Most models agree with each other to within a factor of a few, indicating that the predictions are robust.
Of the lines we consider, \ciii\  (977~\AA) is the strongest line, but it typically traces gas colder than $10^5\,$K. The same is true for \siiv\  (1393,1403~\AA). The second strongest line, \civ\  (1548,1551~\AA), traces circum-galactic gas with $T\sim 10^5\,$K. \ovi\  (1032,1038~\AA) and \neviii\  (770,780~\AA) probe the warmer ($T\sim 10^{5.5}\,$K and $T\sim 10^{6}\,$K, respectively) and more diffuse gas that may be a better tracer of the large scale structure. \nv\  (1239,1243~\AA) emission is intermediate between \civ\ and \ovi. The intensity of all emission lines increases strongly with gas density and metallicity, and for the bright emission it is tightly correlated with the temperature for which the line emissivity is highest. In particular, the \ciii,
\civ, \siiv\ and \ovi\ emission that is sufficiently bright to be potentially detectable in the near future (surface brightness $\ga 10^3$ \phot), comes from relatively dense ($\rho > 10^2 \rho_{\rm mean}$) and metal rich ($Z\ga 0.1 Z_{\sun}$) gas. As such, emission lines are highly biased tracers of the missing baryons and are not an optimal tool to close the baryon budget. However, they do provide a powerful means to detect the gas cooling onto or flowing out of galaxies and groups.
\end{abstract}


\section{Introduction}
\label{intro}

In the high redshift Universe most baryons are believed to reside in gas that is traced by the \lya\ forest which is seen in the spectra of quasars (e.g.\ \citealt{prochaska2008} for a review). In the low redshift Universe, however, the \lya\ forest may trace less than half the baryons. Numerical simulations suggest that this is due to the progressive heating of the diffuse intergalactic medium (IGM) by gravitational shocks produced when structures form (e.g.\ \citealt{sunyaev1972}; \citealt{Nath2001}; \citealt{furlanettoloeb2004}; \citealt{rasera2006}; \citealt{meiksin2009} for a review).

Approximately half of the intergalactic gas is predicted to have temperatures in the range $10^5$ K $<T<10^7$ K at $z\approx 0$ (e.g.\ \citealt{Cen1999}; \citealt{dave2001}; \citealt{bertone2008} for a review). As yet, a large fraction of this shock-heated gas, or warm-hot intergalactic medium (WHIM, hereafter), has not been unambiguously detected.

Metal line transitions in the ultraviolet (UV) band provide a promising route to detecting the cooler fraction of the WHIM with $10^5$ K$<T<10^6$ K. 
The \ovi\ doublet is ideal for tracing this whole temperature range, and a handful of other transitions may help to identify gas in either slightly cooler or warmer temperature regimes. In particular, if collisional ionisation dominates, as we will show to be the case for gas that is dense enough to be detectable in emission, the \civ\ and \nv\ doublets trace gas with $T\sim 10^5$ K, while the \neviii\ doublet is produced by gas with $T\sim 10^6$ K (e.g.\ \citealt{sutherland1993}).
The hottest fraction of the WHIM, with $10^6$ K$<T<10^7$ K, is better traced by \xray\ transitions (e.g.\ \citealt{bregman2007} for a review; \citealt{bertone2010}).

Cooler WHIM gas ($T<10^6$ K) may have been successfully observed in absorption in \fuse\ and \stis\ spectra in the UV band, although the interpretation of these observations remains controversial. Narrow \lya, \civ\ and \ovi\ absorption lines in QSO spectra have been detected in large numbers in high resolution spectra (\citealt{tripp2000}; \citealt{savage2002}; \citealt{danforth2005}; \citealt{tripp2008}; \citealt{danforth2008}; \citealt{cooksey2008}; \citealt{richter2008}; \citealt{thom2008a}; \citealt{thom2008b}; \citealt{wakker2009}; \citealt{cooksey2010}). Intergalactic \neviii, which, so far, has been observed only in the Galactic halo \citep{savage2005}, and \nv\ will also potentially be detected by the Cosmic Origin Spectrograph (\hstcos, \citealt{froning2008}; \citealt{paerels2008}; \citealt{richter2008}) on board the Hubble Space Telescope (HST).

While narrow absorption lines are believed to trace relatively cool and predominantly photo-ionised gas \citep{howk2009}, broad lines have been interpreted as the signature of mostly collisionally ionised gas with temperatures in excess of $10^5$ K (\citealt{richter2004}; \citealt{richter2006}; \citealt{lehner2007}; \citealt{wakker2009}, but see \citealt{oppenheimer2009}). The detection of broad \lya\ absorption lines with velocity widths larger than $b>40$ km s\1 in \fuse\ and \stis\ spectra (\citealt{richter2004}; \citealt{sembach2004}; \citealt{lehner2007}; \citealt{danforth2010}) further improves our understanding of the WHIM and opens the way to accounting for an even larger fraction of the missing baryons and to constrain the sources of the IGM metal enrichment. 
 
While UV absorption lines are a powerful method by which the physical properties of the WHIM can be probed, they only provide 1-dimensional information along the line of sight. Multiple lines of sight from close quasar pairs are necessary to reconstruct the 3-dimensional gas distribution, a challenging task that may be within the reach of \hstcos.
Conversely, emission lines are able to efficiently probe the full 3-dimensional matter distribution by mapping the 2-dimensional distribution of the gas on the sky and by adding information about the third dimension by means of the spectral redshift of the lines (\citealt{furlanetto2004}; \citealt{sembach2009}).
Because emissivity scales as the gas density squared, emission lines are biased towards probing higher density regions than absorption lines.  Therefore, while emission may provide an ideal probe of the properties of the high density WHIM, absorption lines may remain the most powerful tool when investigating mildly dense and underdense regions (see e.g.\ \citealt{furlanetto2004}; \citealt{bertone2008}; \citealt{bertone2010}).

In this work we employ a subset of hydrodynamical, cosmological simulations from the OverWhelmingly Large Simulations (\owls, hereafter) project \citep{schaye2010} to predict the intensity of a sample of UV emission lines that could potentially be detected by current and future instruments, such as the Faint Intergalactic Redshifted Emission Balloon (\fireball, \citealt{tuttle2008}) and the Advanced Technology Large Aperture Space Telescope\footnote{http://www.stsci.edu/institute/atlast} (\atlast, \citealt{postman2008}).
In a companion paper (\citealt{bertone2010}, Paper I in the following) we have investigated the soft \xray\ emission from the WHIM using the same methodology, and we refer to that work for more details about the numerical aspects of our calculations.

The \owls\ runs are ideal for this study because they couple a large simulated dynamical range at high resolution with a large set of models in which the implementation of several physical processes has been varied. This feature in particular allows us to investigate the dependence of the predicted UV emission on a number of uncertain physical prescriptions, in addition to considering standard tests for mass resolution and the size of the simulation box. In this work we use simulations of 100 \hm\ comoving Mpc on a side and focus on the strongest line of the UV doublets \civ, \siiv, \nv, \ovi\ and \neviii\ and the singlet \ciii, listed in Table \ref{eltable}. The intensity of the weaker line in the doublets can be recovered by multiplying the intensities derived in this study by a factor of 0.5, which corresponds to the difference in the line emissivities in the doublet.

\begin{table}
\centering
\caption{List of emission lines. The first column shows the ion giving rise to the line, the second and the third columns the wavelengths of the two components of the doublet. The strongest component is listed in column 2 and its energy in column 4. \ciii\ is the only singlet in the sample.}
\begin{tabular}{l r@{}l r@{}l r@{}l}
\hline
\hline
Ion & $\lambda_{1}$ \; & (\AA) & $\lambda_{2}$ \; & (\AA) & $E_{1}$ \; & (eV) \\
\hline
\ciii   & 977.  & 03  & $-$   &     & 12. & 690 \\
\civ    & 1548. & 187 & 1550. & 774 & 8.  & 008 \\
\siiv   & 1393. & 755 & 1402. & 770 & 8.  & 896 \\
\nv     & 1238. & 821 & 1242. & 804 & 10. & 008 \\
\ovi    & 1031. & 912 & 1037. & 613 & 12. & 015 \\
\neviii & 770.  & 409 & 780. & 324 & 16. & 094 \\
\hline
\hline
\end{tabular}
\label{eltable}
\end{table}

This paper is organised as follows. We briefly describe the simulations and the method by which we calculate the gas metal line emission in Section \ref{owls}. Section \ref{emission} contains our results for $z=0.25$, while in Section \ref{angle} we test the effect of varying the angular resolution used to build the maps. In Section \ref{redshift} we consider the redshift dependence of the flux intensity. Section \ref{em_from} investigates what kind of gas produces most of the emission and Section \ref{em_model} examines the dependence of the results on the physical model. We compare our results to those of \citet{furlanetto2004} in Section~\ref{furla} and we discuss the prospects for observing the UV emission with future space-borne telescopes in Section \ref{uv_em}. Finally, we summarise our conclusions in Section \ref{summary} and describe numerical convergence tests in the Appendix.

Throughout this paper we will assume a WMAP3 $\Lambda$CDM cosmology \citep{spergel2007} with parameters $\Omega_{\rm m}=0.238$, $\Omega_{\rm b}=0.0418$, $\Omega_\Lambda=0.762$, $n=0.951$, and $\sigma_8=0.74$. The Hubble constant is parametrised as $H_{\rm 0} = 100$ $h^{-1}$ km s$^{-1}$ Mpc$^{-1}$, with $h=0.73$.

\section{Numerical methods}
\label{numbers}

In Section \ref{owls} we briefly describe the subset of simulations from the \owls\ project employed in this study, while in Sections \ref{tables} and \ref{maps} we review the methods by which we calculate emission line intensities, and generate synthetic surface brightness maps, respectively.

\subsection{Numerical simulations}
\label{owls}

In this work we employ a subset of the cosmological, hydrodynamical simulations that comprise the \owls\ project \citep{schaye2010}. A detailed description of the numerical methods employed in the OWLS project is given in \citet{schaye2010}. Here we give a brief overview of the subgrid physical modules used in the reference model (\default, hereafter) that is used to derive most of the results in this paper. In Section \ref{em_model} we discuss a number of other models that incorporate variations of the physical modules, implemented one at a time.

The simulations were run using a significantly extended version of the parallel PMTree-SPH code {\sc gadget III} \citep{springel2005g}. Here we use runs with box sizes 100 \hm\ comoving Mpc containing $512^3$ particles of both gas and dark matter. The baryonic particle mass is $8.66\times 10^7$ \hm\  \msun.

Gas cooling is implemented following the prescription of \cite{wiersma2009a}\footnote{Using their equation (3) rather than (4) and {\sc cloudy} version 05.07 rather than 07.02, as used in that paper.}. Net radiative cooling rates are computed and tabulated element by element for all 11 elements (H, He, C, N, O, Ne, Mg, Si, S, Ca and Fe) tracked by the simulation in the presence of the cosmic microwave background and the \citet{haardt2001} model for the UV/X-ray background radiation from quasars and galaxies. The tables, created with the publicly available photo-ionization package \cloudy\  (last described by \citealt{ferland1998}) assume the gas to be optically thin and in (photo-)ionization equilibrium, are interpolated over density, temperature and redshift.

Star formation is implemented following the sub-grid prescription of \citet{schaye2008}.  Gas with densities exceeding the threshold for the onset of thermo-gravitational instability ($n_{\rm H} = 0.1$ cm\3; \citealt{schaye2004}) is considered to be multiphase and follows an effective equation of state of the form $P\propto \rho^{\gamma_{\rm eff}}$, normalised to $P/k=1.08\times 10^3$\, K cm\3\ at the density threshold. Multiphase gas forms stars at a pressure-dependent rate that reproduces the observed Kennicutt-Schmidt law \citep{kennicutt1998}. The prescription by which the simulations track the subsequent chemical enrichment of the gas is described in \citet{wiersma2009b}. In brief, we follow the timed release of 11 different elements from massive stars (Type II supernovae and stellar winds) and intermediate mass stars (Type Ia supernovae and asymptotic giant branch stars), assuming a Chabrier \citep{chabrier2003} IMF in the mass range 0.1-100\,\msun.

Energy injection from supernovae is included as kinetic feedback following the prescription of \citet{vecchia2008}. Two parameters are used to describe the energy injection: the wind velocity, $v_{\rm w}$, and the mass loss rate, $\eta=\dot{m}_{\rm w}/\dot{m}_\ast$, which describes the amount of gas kicked into the wind, $\dot{m}_{\rm w}$, as a function of the rate of star-formation, $\dot{m}_\ast$. The \default\ model uses $\eta=2$ and $v_{\rm w}=600$ \kms, which correspond to 40\% of the supernova energy being available in the form of kinetic energy.  The \default\ model does not include a prescription for AGN feedback, but we will consider a model with AGN in Section \ref{em_model}.

\subsection{The emissivity tables}
\label{tables}

\begin{figure*}
\centering
\includegraphics[width=\textwidth]{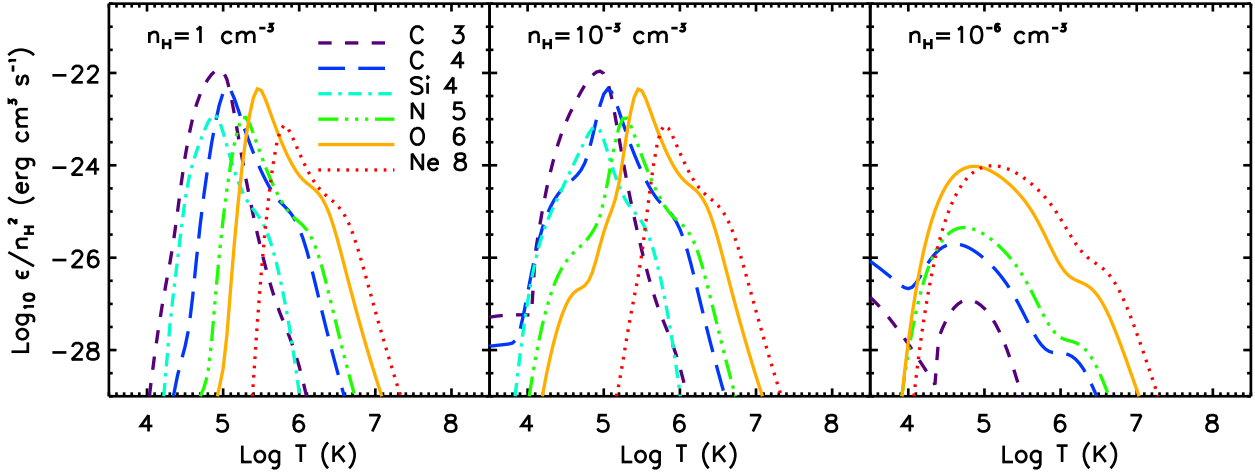}
\caption{The normalised emissivity, $\epsilon / n_{\rm H}^{2}$, in units of erg~cm$^3\,$s\1, of a selection of UV emission lines as a function of temperature assuming solar abundances at $z=0.27$. The left, middle, and right panels show results for constant hydrogen number densities $n_{\rm H} = 1$, $10^{-3}$ and $10^{-6}\,\cm^{-3}$, respectively. If the gas is collisionally ionised, the normalised emissivity is independent of the gas density. Cool diffuse gas is mostly photo-ionised, while collisional ionisation equilibrium dominates for the highest temperatures and densities.}
\label{lines}
\end{figure*}

\begin{table}
\centering
\caption{Adopted solar abundances, from \citet{allende2001}, \citet{allende2002} and \citet{holweger2001}.}

\begin{tabular}{l c | l c}
\hline
\hline
Element & $n_i / n_{\rm H}$ & Element & $n_i / n_{\rm H}$ \\
\hline
H  & 1                    & Mg & 3.47$\times 10^{-5}$ \\
He & 0.1                  & Si & 3.47$\times 10^{-5}$ \\
C  & 2.46$\times 10^{-4}$ & S  & 1.86$\times 10^{-5}$ \\
N  & 8.51$\times 10^{-5}$ & Ca & 2.29$\times 10^{-6}$ \\
O  & 4.90$\times 10^{-4}$ & Fe & 2.82$\times 10^{-4}$ \\
Ne & 1.00$\times 10^{-4}$ &    &                      \\
\hline
\hline
\end{tabular}
\label{table_abund}
\end{table}

The intensity of the emission lines is calculated following the procedures of Paper I. We generated tables of the line emissivity as a function of hydrogen number density, temperature and redshift for about 2000 of the strongest lines of the 11 elements tracked by \owls. The line emissivities are calculated using \cloudy\  (version c07.02.02, \citealt{ferland1998}, released in July 2008) for an optically thin gas in ionisation equilibrium in the presence of the cosmic microwave background and a uniform, evolving meta-galactic UV radiation field \citep{haardt2001} from galaxies and quasars. The same assumptions were used in the calculation of the radiative cooling rates \citep{wiersma2009a}.
The tables sample the temperature range $10^2$ K $< T < 10^{8.5}$ K in bins of $\Delta {\rm Log}_{10}T=0.05$ and the hydrogen number density range $10^{-8}$ cm\3\ $< n_{\rm H} < 10$ cm\3\ in bins of $\Delta {\rm Log}_{10} n_{\rm H}=0.2$.

The tables were computed for solar metallicity, with the adopted solar abundance $Z_{\sun} =0.0127$. This corresponds to the value obtained using the default abundance set of \cloudy, listed in Table \ref{table_abund}, which is a combination of the abundances of \citet{allende2001}, \citet{allende2002} and \citet{holweger2001}. Our adopted abundance set may differ significantly from that estimated by \citet{lodders2003}. In \cloudy, the oxygen abundance in particular is about 20 per cent lower than estimated by \citet{lodders2003}.
The differences in the assumed element abundances should be kept in mind when comparing with predictions from different codes.

Fig. \ref{lines} shows the normalised emissivities of the UV lines listed in Table \ref{eltable} as a function of temperature, for constant hydrogen number densities $n_{\rm H} = 1$, $10^{-3}$ and $10^{-6}$ cm\3\  (left, middle and right panels, respectively). Fig. \ref{lines} demonstrates that collisional ionisation dominates at the highest temperatures and densities, while photo-ionisation becomes the main ionisation process only when both the density and the temperature are very low. This can be seen by comparing the three panels of Fig. \ref{lines} and noting that the normalised emissivity peaks at a well defined temperature and is independent of the density, if collisional ionisation dominates. As we will see in the following, most of the gas emission is produced in collisionally ionised regions with gas densities $n_{\rm H}\gtrsim 10^{-3}$ cm\3.

The assumption of ionisation equilibrium, also used to calculate the cooling tables \citep{wiersma2009a}, is justified for photo-ionised regions and
dense gas in the centres of clusters (see \citealt{bertone2008} for a review). However, non-equilibrium processes may become important in the WHIM and in the outer regions of groups and clusters (\citealt{hughes1994}; \citealt{Yoshida2005}; \citealt{Yoshikawa2006}; \citealt{CenFang2006}; \citealt{gnat2007}). \citet{gnat2009}, in particular, have recently shown that non-equilibrium processes affect the cooling of shock-heated gas with temperatures $10^4\,\K <T<10^7\,\K$, and that the amplitude of the deviations from ionisation equilibrium increases with the gas metallicity. On the other hand, \citet{Yoshikawa2006} found the effect of non-equilibrium ionisation to be small for WHIM emission. Finally, we note that because these studies ignored photo-ionisation, they may have overestimated the importance of non-equilibrium ionisation.

\subsection{The metal line emission}
\label{maps}

The emission of the sample of UV lines listed in Table \ref{eltable} is calculated following the procedures described in Paper I and we refer the interested reader to that work for the full details.

We calculate emission only for non-star forming gas with $n_{\rm H} < 0.1$ cm\3\ because our simulations lack the resolution and the physics to model the emission from higher density gas in the interstellar medium, which is expected to be multiphase \citep[see the discussion in][]{schaye2008}.

The luminosity of emission line $l$ for gas particle $i$, in units of erg s\1, is calculated as a function of its gas mass $M_{{\rm gas},i}$, density $\rho_i$, hydrogen number density $n_{{\rm H},i}$ and element abundance $X_{{\rm y},i}$ as
\begin{equation}\label{lumin}
L_{i,l}\left(z\right) = \varepsilon_{i,l,\odot} \left( z, T_{\rm i},n_{\rm H} \right) \frac{M_{{\rm gas},i}}{\rho_i} \frac{X_{{\rm y},i}}{X_{\rm y \sun}},
\end{equation}
where $\varepsilon_{i,l,\odot} \left( z, T_{\rm i},n_{\rm H} \right)$ the emissivity of the line $l$, in units of erg cm\3\ s\1, bi-linearly interpolated (in logarithmic space) from the emissivity tables as a function of the particle temperature ${\rm Log}_{10} T_{\rm i}$ and hydrogen number density ${\rm Log}_{10} n_{{\rm H},i}$ at the desired redshift. $X_{{\rm y},i}$ is the mass fraction of element $y$ and $X_{\rm y \sun}$ the corresponding solar value. We use the ``smoothed'' element abundances described in \citet{wiersma2009b}, which alleviates the effect of the lack of metal mixing inherent to SPH simulations, although it does not solve the problem. The main effect of the use of smoothed abundances is to spread the emission over slightly greater gas masses, particularly in regions with lower metallicities. 

The corresponding particle flux $F_{i,l}$, in units of photon s\1\ cm\2, is given by
\begin{equation}
F_{i,l} = \frac{L_{i,l}}{4\pi D_{\rm L}^2} \frac{\lambda_l}{h_{\rm P}c} \left(1+z\right),
\end{equation}
where $D_{\rm L}$ the luminosity distance, $h_{\rm P}$ the Planck constant, $c$ the speed of light and $\lambda_l$ the rest-frame wavelength of the emission line $l$.

To create maps of the line emission, the particle fluxes are projected on to a 2-dimensional grid using a flux-conserving smoothed particle hydrodynamics interpolation scheme. Finally, surface brightness ($S_{\rm B}$, hereafter) maps are computed by dividing the flux in each pixel, $p$, by the solid angle $\Omega$ it subtends: $S_{{\rm B}p,l} = \sum_i F_{i,l} / \Omega$.

The number of pixels in the grid is determined by the angular resolution required to match the capabilities of current or future instruments. We investigate how fluxes change with grid angular resolution in Section \ref{angle}. A region with comoving size 100 \hm\ Mpc at $z=0.25$ corresponds to a field of view of about 8 degrees for our chosen cosmology. An angular resolution of 15" on the sky, used throughout this work unless otherwise noted, requires a grid with $1926^2$ pixels. At $z=0.25$ this angular resolution corresponds to a physical scale of 42 \hm\ kpc, which is about 21 times larger than the maximum gravitational softening scale used in the simulation \citep{schaye2010}.
We divide each simulated box of 100 \hm\ Mpc into 5 slices, each 20 \hm\ Mpc thick, and create as many emission maps, which we combine to calculate the flux probability distribution function (PDF, hereafter) corresponding to this angular resolution and slice thickness. At $z=0.25$, a slice thickness of 20 \hm\ Mpc corresponds to an energy resolution $\Delta E / E \sim 6 \times 10^{-3}$, which is roughly comparable to that of \fireball.

\section{Maps of the UV emission}
\label{emission}

\begin{figure*}[Ht]
\centering
\includegraphics[width=\textwidth]{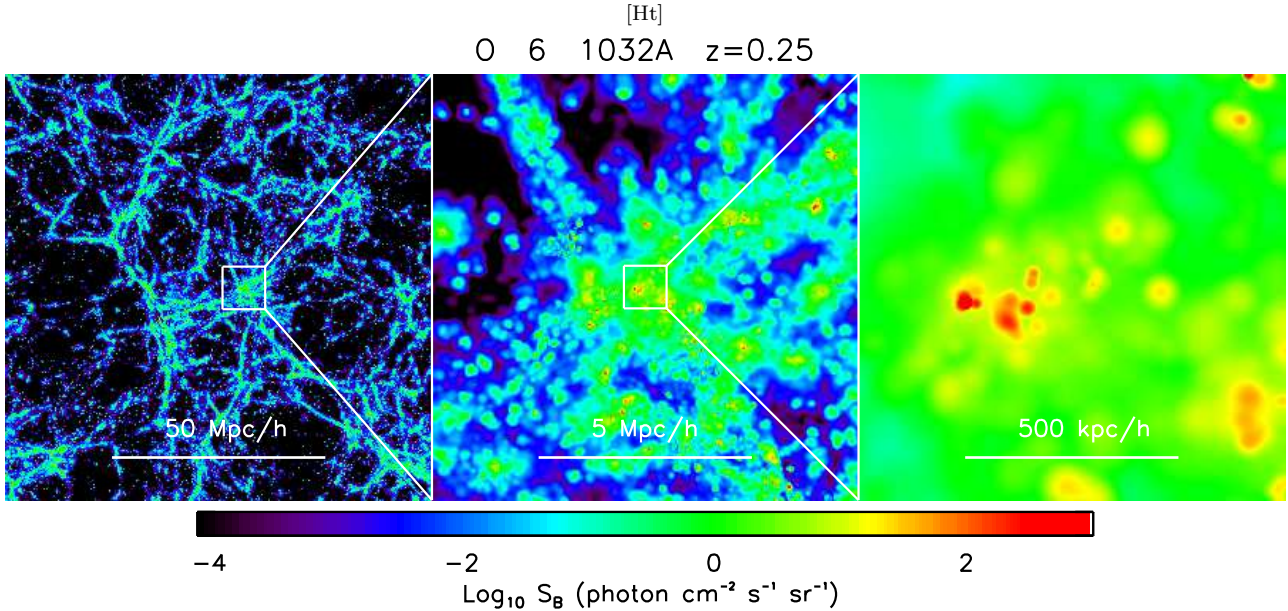}
\caption{Zoom into an \ovi\ 1032 \AA\ emission map of a high-density region at $z=0.25$. The object at the centre of the displayed region is a relatively small group. The slice thickness is 20 \hm\ comoving Mpc (which corresponds to $1772$~\kms\ for all panels). From left-to-right, the pixel sizes are 100", 10" and 1" and the angular sizes of the regions shown are 8 degrees, 48' and 4.8' (corresponding to 100, 10, and 1~\hm\ comoving Mpc), respectively. The blobs in the right panel are resolved substructures.}
\label{zoom_o6}
\end{figure*}

\begin{figure*}
\centering
\includegraphics[width=\textwidth]{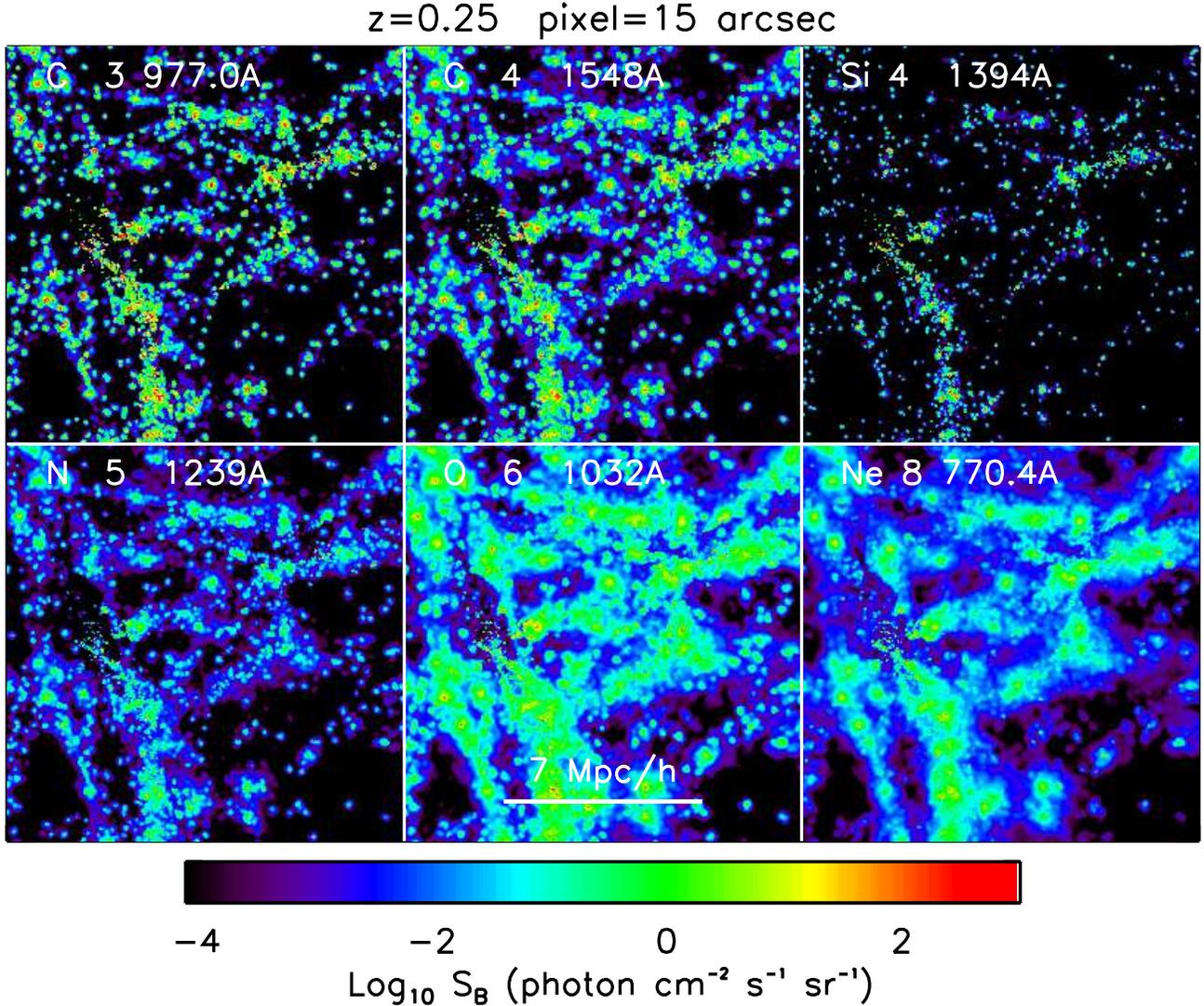}
\caption{Maps of the emission for a sample of six different UV lines, listed in table \ref{eltable}, at $z=0.25$. The figure shows a zoom of a region, with a comoving size of 14 \hm\ Mpc (1.12 degrees on the sky), that includes the largest group in the simulation, with a total mass of a few times $10^{14}$ M$_{\sun}$. The density, temperature, and metallicity in this region are shown in Fig. \ref{dtz}. All panels assume the same angular resolution (15", which corresponds to a physical size of 42~\hm\ kpc) and slice thickness (20~\hm~comoving Mpc, which corresponds to 1772~\kms). The maximum of the colour scale has been set to a fraction of the real maximum to enhance the emission of the weaker lines and of low-density regions.
While strong \civ\ emission is concentrated around the haloes of galaxies, \ovi\ and \neviii\ lines trace more diffuse gas in a variety of structures such as filaments and haloes. Emission from UV lines is suppressed at the centres of large haloes, where the gas is hotter than $10^6$ K and \xray\ emission dominates.}
\label{uv_maps}
\end{figure*}

\begin{figure*}
\centering
\includegraphics[height=\textwidth,angle=90]{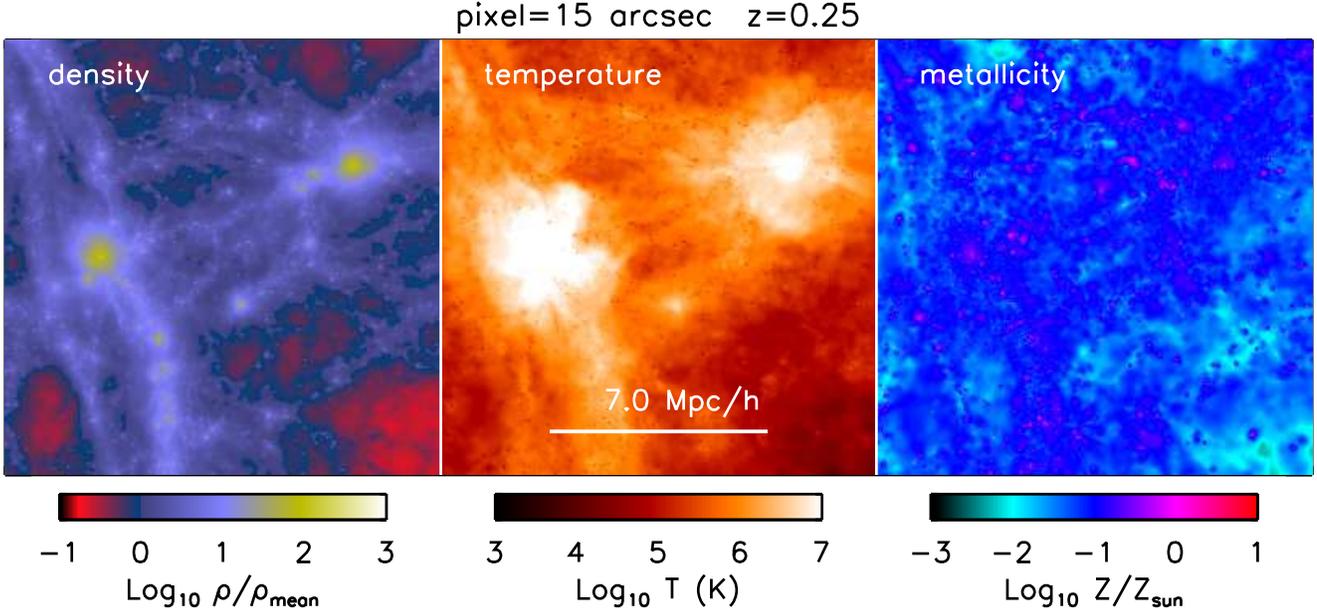}
\caption{As Fig.~\protect\ref{uv_maps}, but showing maps of the gas density (left panel), temperature (middle panel) and metallicity (right panels) for both star forming and non-star forming gas.}
\label{dtz}
\end{figure*}

\begin{figure}
\centering
\includegraphics[width=8.4cm]{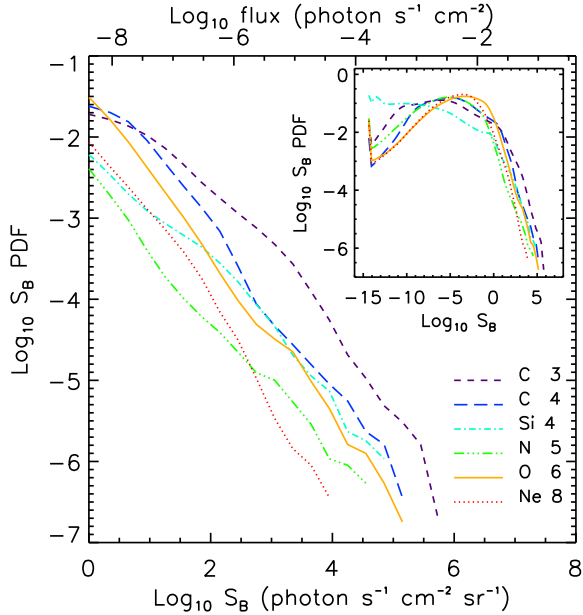}
\caption{The surface brightness PDFs of our sample of UV emission lines, also shown in Fig. \ref{uv_maps} and listed in Table \ref{eltable}, at $z=0.25$.
The main plotting area shows only the high flux tail of the distributions, while the full PDFs are shown in the inset. The upper axis shows the flux per pixel in units of photon s\1\ cm\2. The pixel size is 15" (which corresponds to a physical size of 42~\hm~kpc) and PDFs are calculated using five slices through the simulation box that are each 20~\hm~comoving Mpc thick (which corresponds to 1772~\kms). In Appendix~\ref{thick} we show that for sufficiently large fluxes ($>10^{-2}$~\phot\ for \ovi) the PDF is proportional to the slice thickness.
The \ciii\ and \civ\ lines are the strongest emission lines in the sample, closely followed by \ovi. \nv\ and \neviii\ fluxes are weaker than \civ\ by about an order of magnitude.}
\label{uv_pdf}
\end{figure}

\begin{figure}
\centering
\includegraphics[width=8.4cm]{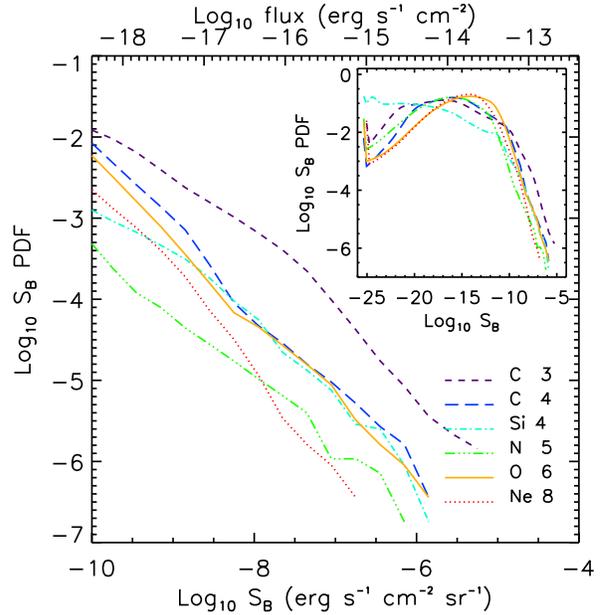}
\caption{As Fig. \ref{uv_pdf}, but with the emission line flux expressed in erg s\1\ cm\2\ sr\1, instead of \phot. The relative strength of the lines varies slightly when converted to these units and is more easily comparable to the flux limits proposed for future missions.}
\label{uv_pdf2}
\end{figure}

\begin{figure*}
\centering
\includegraphics[width=0.9\textwidth]{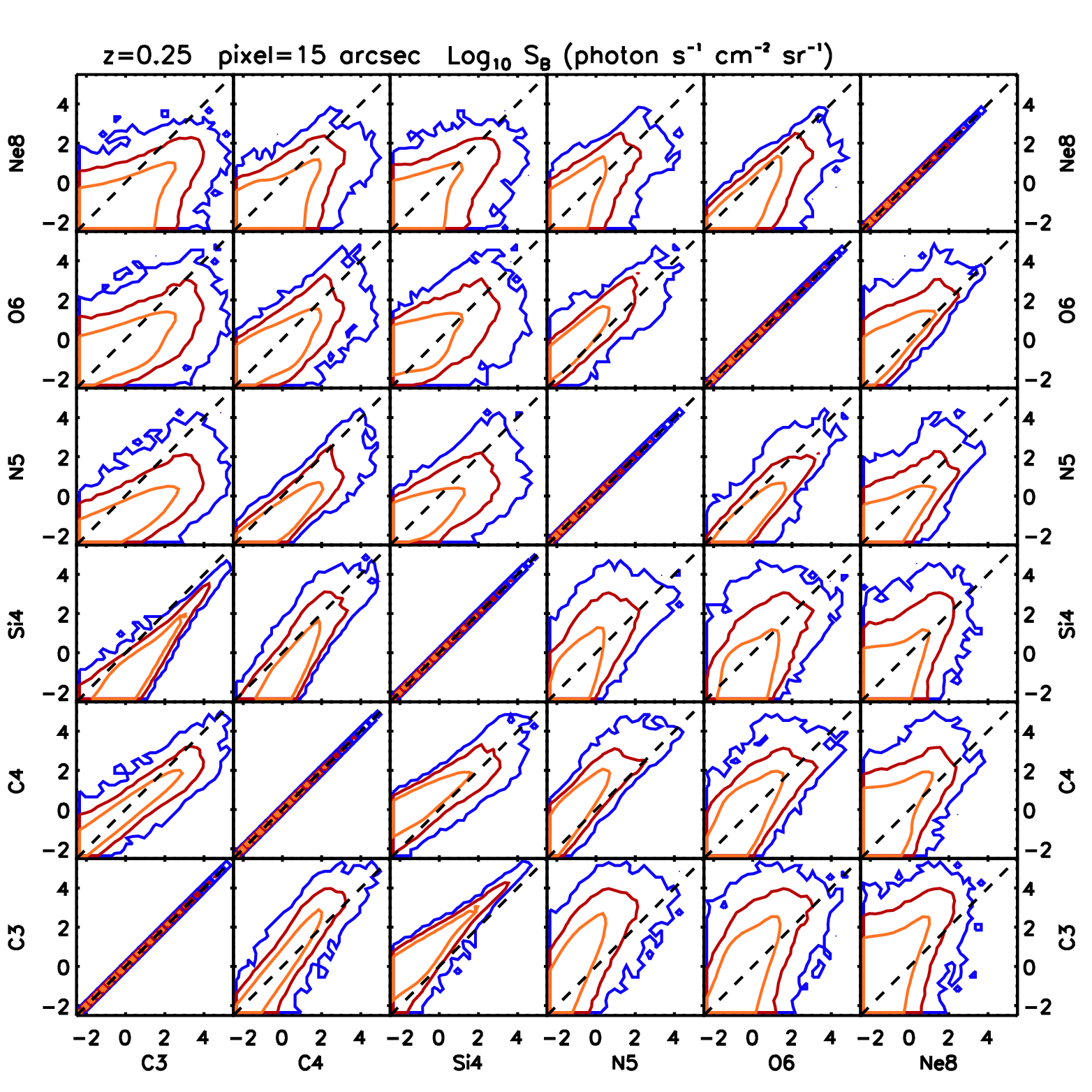}
\caption{Comparison of the intensities of the lines listed in Table~\ref{eltable}. The contours show the number density of pixels and are logarithmically spaced by 1.5~dex. The distributions have been calculated using five 20 \hm\ comoving Mpc thick slices through the simulation box at $z=0.25$, assuming an angular resolution of 15". Both axes show the line surface brightness ${\rm Log}_{10} S_{\rm B}$ in units of \phot. The dashed diagonal line in each panel indicates equal fluxes. The \ciii\ and \civ\ lines produce 
the highest fluxes. Lines whose emissivities peak at similar temperatures (i.e.\ panels near the diagonal running from the bottom-left to the top-right of the plot) are strongly correlated.} 
\label{fluxzz}
\end{figure*}

In this Section we present emission maps for the strongest and most interesting UV emission lines we have identified as tracers of the cooler WHIM component with $T\la 10^6$ K.

Fig. \ref{zoom_o6} gives an impression of the morphology of the UV emission from large-scale structure. The left panel depicts the full simulation box and the two companion panels show two consecutive zooms into the core of a galaxy group. The displayed regions decrease in size by a factor of ten, ranging from 100 \hm\ Mpc to 1 \hm\ Mpc.

Fig. \ref{uv_maps} shows surface brightness maps at $z=0.25$ for the six UV lines considered in this work. The surface brightness PDFs are presented in Fig. \ref{uv_pdf} in units of \phot, and in Fig.~\ref{uv_pdf2} in units of erg s\1\ cm\2\ sr\1. For comparison, the density, temperature, and metallicity in this region are shown in Fig. \ref{dtz}, which is the same as Fig.~13 of Paper I. This is the same region of space shown in all the emission maps in this paper, with the exception of Fig. \ref{zoom_o6}.

Fig. \ref{uv_maps} reveals that none of the UV lines gives rise to much emission from the largest haloes (cf. Fig. \ref{dtz}) as in these haloes gas is shock heated to $T>>10^6$ K, far above the temperatures at which the ions responsible for UV emission are abundant. The \ovi\ and \neviii\ fluxes show similar spatial distributions with most of the emission associated with filaments. The \ciii, \civ\ and \siiv\ emission maps demonstrate a much more clumpy morphology and their emission is associated chiefly with circum-galactic gas. This is a direct consequence of the fact that the peak of the emissivity shifts to higher temperature with increasing atomic number and ionization state, as can be seen from Fig. \ref{lines}.

By comparing these results to those of Paper I, we find that the UV emission has a different spatial distribution from the soft X-ray emission. While the \xray\ emission mostly comes from the warmest gas in the largest haloes and filaments, UV lines trace smaller structures, such as galaxies and small groups.
This difference can be partially explained by the different peak emissivity temperatures of the different ions.
Here we find that \ovi\ and \neviii\ lines trace primarily the highest temperature, shocked gas ($\sim10^6\,$K) in filaments, in the surroundings of groups and in the intragroup medium itself, whereas \ciii, \civ\ and \siiv\ lines trace gas at lower temperatures, have very inhomogeneous distributions and trace gas associated with galaxies. \siiv\ lines trace only very compact regions of space, that can be identified with gas in the haloes of galaxies. The spatial distribution of \ciii\ emission is intermediate between those of \siiv\ and \civ.

The PDFs in Figs. \ref{uv_pdf} and \ref{uv_pdf2} demonstrate that \ciii\ is the strongest of the emission lines considered in this study, and the strongest fluxes are concentrated in very few pixels (or equivalently, in very small regions). This is partly a consequence of the \ciii\ line emissivity (Fig. \ref{lines}) which, at high density, has a peak value at $T\sim 10^5$ K that is larger than that of all other UV lines by 0.5 dex, and partly of the large amount of metals in the C$^{2+}$ phase, as we show in Section \ref{em_plane}. \civ\ surface brightnesses are typically an order of magnitude lower than those of \ciii, in high density regions.
\nv\ emission is weaker than \civ\ by a factor of 10-100 and \neviii\ by a factor of about 10. On average, \ovi\ emission is fainter than \civ\ by a factor of about 2. The different spatial distributions of the ions that trace high temperature gas (\ovi, \neviii, \nv) and those that trace cooler gas (\ciii, \civ, \siiv) mean that in spite of not being the strongest line in high density regions, \ovi\ emission is able to exceed the emission from all other lines in low density regions, where carbon and silicon emission declines steeply.  We therefore conclude that the best line in the UV band for probing the properties of the diffuse WHIM is \ovi, whereas the strong lines such as \ciii\ and \civ\ that trace cooler gas offer the exciting opportunity of observationally probing the colder gas around, and accreting onto, galaxies.

Fig.~\ref{fluxzz} shows the relative intensity of lines with respect to each other. Where contours lie above the dashed diagonal line, the emission line on the $y$-axis produces the highest flux, when below, the line on the $x$-axis is strongest. Line pairs with similar peak temperatures produce narrow contours that may lie on either side of the diagonal, while line pairs with different peak temperatures produce broad profiles. The large dispersion indicates that different lines are dominant in different density and temperature regimes.
Fig.~\ref{fluxzz} illustrates the distribution of line ratios that can be 
expected when surveying the WHIM emission in a large field and may ultimately help to identify lines in different regions.

\section{Angular resolution}
\label{angle}

\begin{figure}
\centering
\includegraphics[width=8.4cm]{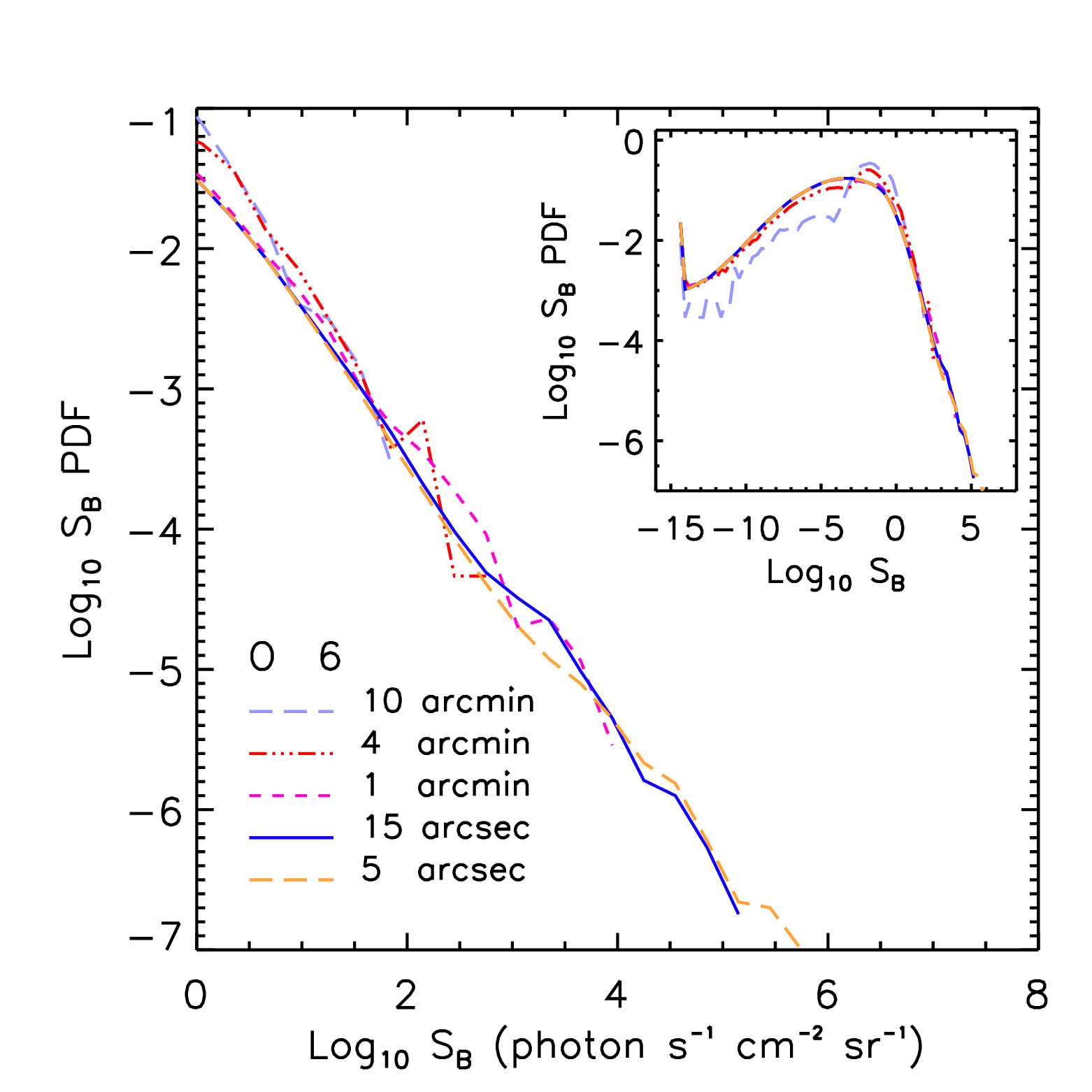}
\caption{As Fig.~\protect\ref{uv_pdf}, but for \ovi\ emission as a function of the angular resolution of the emission maps. The pixel sizes are $\vartheta = 5"$, $15"$, $1'$, $4'$, and $10'$, corresponding to physical sizes of 14 \hm\ kpc, 42 \hm\ kpc, 166 \hm\ kpc, 665 \hm\ kpc and 1.661 \hm\ Mpc, respectively. 
With the exception of the $\vartheta=10'$ and 4' cases, the results are converged at intermediate and low surface brightness ($S_{\rm B}<10^2$ \phot). The maximum predicted surface brightness increases steadily with angular resolution, a consequence of the fact that very strong emission on scales smaller than the resolution is smoothed out. There is no clear convergence at low angular resolutions ($\vartheta >1'$).}
\label{angle_figure}
\end{figure}

In this Section we test the effect of varying the angular resolution used to make the emission maps. Fig. \ref{angle_figure} compares the PDFs of the surface brightness in maps with resolutions $\vartheta = 5", 15", 1', 4'$ and $10'$, which correspond to physical scales varying from 14 \hm\ kpc to 1.661 \hm\ Mpc.

We find that most distributions, with the exception of the $\vartheta = 10'$ case, converge well at intermediate and low surface brightness. We find no convergence at the bright end, where the highest predicted surface brightness steadily increases with increasing resolution, at least on the scales considered here. This is a consequence of the fact that large pixel sizes prevent the detection of very strong emission coming from structures with an angular size smaller than that used to observe them.
This said, diffuse gas on scales larger than galactic scales, such as the WHIM gas, can be observed with an angular resolution of about 10-15" at $z\sim 0.25$ without significant loss of information. Higher angular resolution, however, might be desirable when observing gas with more clumpy spatial distribution on smaller scales, such as outflowing gas in galactic winds.
For reference, in the \owls\ cosmology a pixel size of 15" subtends a comoving (proper) size of about 52 (42) \hm\ kpc at $z=0.25$, 98 (66) \hm\ kpc at $z=0.5$ and 175 (88) \hm\ kpc at $z=1$.

\section{Evolution}
\label{redshift}

\begin{figure}
\centering
\includegraphics[height=8.4cm]{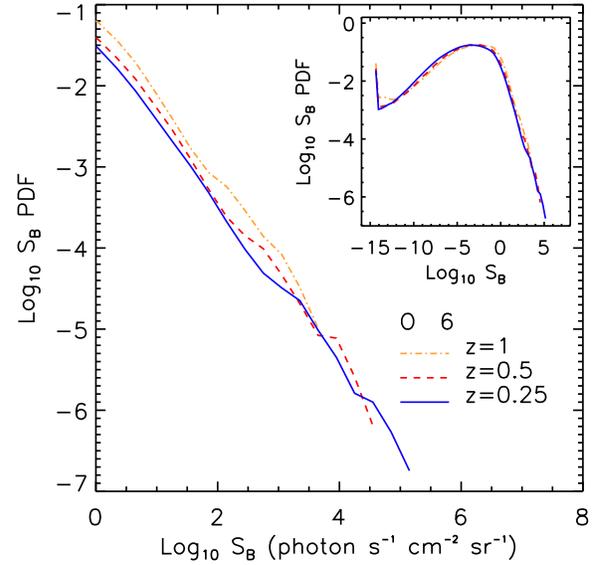}
\caption{As Fig. \ref{uv_pdf}, but for \ovi\ emission as a function of redshift. The surface brightness PDFs are similar at low fluxes, but diverge at the highest fluxes. In particular, the highest fluxes are found at the lowest redshift, which may be mostly due to the varying physical size of the pixels.}
\label{redshift_pdf}
\end{figure}

In this Section we briefly discuss how the intensity of the emitted flux varies with cosmic time.

In Fig. \ref{redshift_pdf} we show the surface brightness PDF of \ovi\ emission at $z=0.25, 0.5$ and 1.
The overall shape of the PDF does not vary substantially, but the highest predicted flux increases with decreasing redshift. This is at least partly due to the fact that at the constant angular resolution of 15" assumed for the maps, the physical size that corresponds to a pixel increases steadily with redshift, from 42 \hm\ kpc at $z=0.25$ to 88 \hm\ kpc at $z=1$. As shown in Section \ref{angle}, the maximum flux decreases with increasing pixel size.
However, the emissivity also depends on the density, temperature and metallicity of the emitting gas, all of which evolve with cosmic time.

\section{What type of gas dominates the emission?}
\label{em_from}

In this Section we investigate which gas contributes most of the UV emission. In particular, we study the dependence of the emission on density and temperature in Section \ref{em_plane}. We present the correlations between the observed fluxes and the physical properties of the gas in Section \ref{correlations}, and we finally calculate the emission from gas in different density and temperature ranges in Section \ref{cuts}.

\subsection{Temperature-density plane}
\label{em_plane}

\begin{figure*}
\hspace{-1cm}
\includegraphics[width=0.8\textwidth]{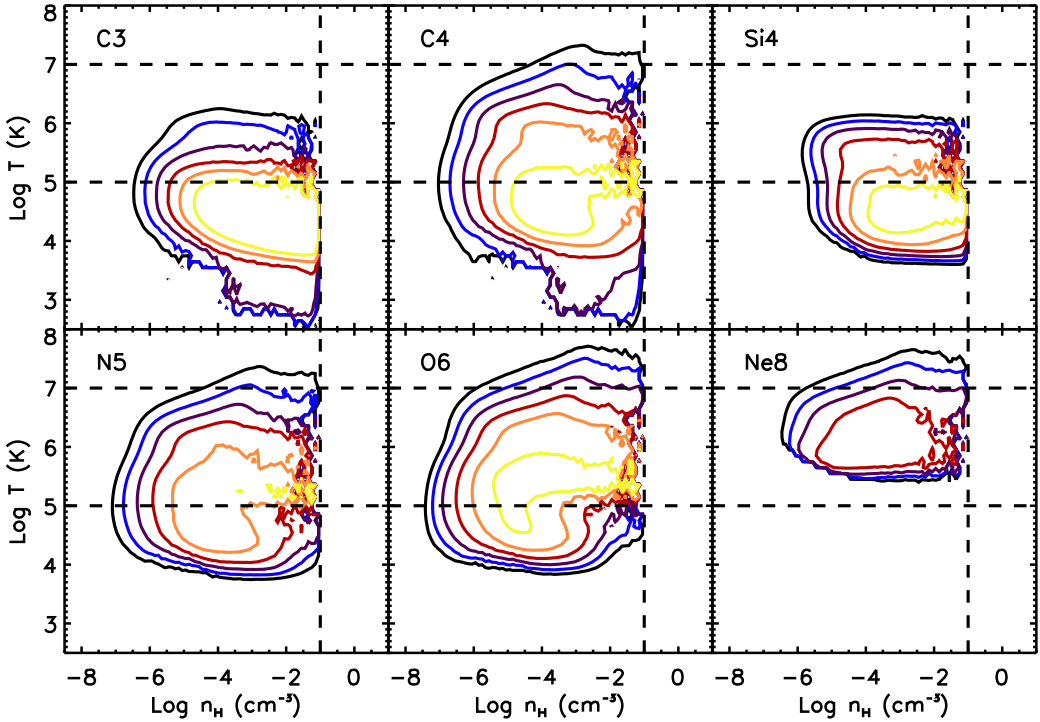}
\caption{The UV emission-weighted temperature-density distribution at $z=0.25$. The distribution is weighted by the emission of a different ion, as indicated in the top left corner of each panel. The ions we consider here are \ciii, \civ, \siiv, \nv, \ovi\ and \neviii. The contours are logarithmically spaced by 1.5 dex and represent equal emission levels. The horizontal lines at constant temperature highlight the WHIM range, while the vertical line at $n_{\rm H}=0.1$ cm\3\ indicates the limit above which we impose an effective equation of state on to star-forming particles. For reference, the cosmic mean density corresponds to $n_{\rm H}\sim 10^{-7}$ cm\3.
\civ\ emission mostly traces dense warm gas in the haloes of galaxies. The bulk of the emission from \ovi\ arises from equally dense, but hotter gas, while \nv\ emission traces an intermediate phase of gas between those of \civ\ and \ovi. \neviii\ emission is produced by hot gas with $T\ga 10^6$ K, whose distribution partially overlaps with that of \ovi, but is only marginally traced by \civ\ lines.}
\label{uvcont}
\end{figure*}

In Paper I we demonstrated that at $z=0.25$ the bulk of the metals do not trace the bulk of the IGM mass.
In fact, most gas mass resides in the low density IGM, while metals are more common in the moderately dense, WHIM gas (see also \citealt{wiersma2009b}) and in the relatively dense gas with $n_{\rm H} > 0.01$ cm\3\ and $T<10^5$ K that is cooling on to galaxies. As stated in Section \ref{maps}, we neglect the emission from the interstellar medium, assumed here as all the gas with $n_{\rm H} > 0.1$ cm\3, because the simulations lack the resolution and the physics to model its multi-phase composition.
We found in Paper I that the \xray\ emission traces neither the mass nor the metals in the IGM, but the densest and most metal enriched phases with temperatures close to the peak temperatures of the emissivity curves.

In Fig. \ref{uvcont} we show the emission-weighted temperature-density
distributions of the gas for our sample of UV lines, where the weight is given by the emission of a different line in each panel.
The vertical line at $n_{\rm H}=0.1$ cm\3\ indicates the density threshold above which star forming particles obey an effective equation of state. The horizontal lines at $T=10^5$ K and $10^7$ K delimit the temperature range of the WHIM.
We find that most of the emission traces moderately overdense gas with $n_{\rm H}>10^{-5}$ cm\3\ and that the bulk of the emission clusters around the emissivity peak temperature of each line. As such, \neviii\ is a good tracer of hot gas with $T\sim 10^6$ K, while \nv\ and \ovi\ trace the cooler fraction of the WHIM with temperatures in the range $10^5$ K$<T<10^6$ K. \civ\ traces gas slightly cooler than the WHIM, with $10^4$ K$<T\la 10^{5.5}$ K. The bulk of the \ciii\ and \siiv\ emission arises from gas with $10^4$ K$<T\la 10^{5}$ K. The large scatter in some of the distributions shows that small fractions of the \civ, \nv\ and \ovi\ emission trace gas with temperatures that can differ by more than an order of magnitude from the peak temperatures of the emissivity curves, while the relatively narrow distributions in temperature of the \ciii\ and \siiv\ emissions indicates very inefficient emission outside that narrow temperature range.

\subsection{Correlations between luminosity and gas properties}
\label{correlations}

\begin{figure*}
\centering
\includegraphics[width=\textwidth]{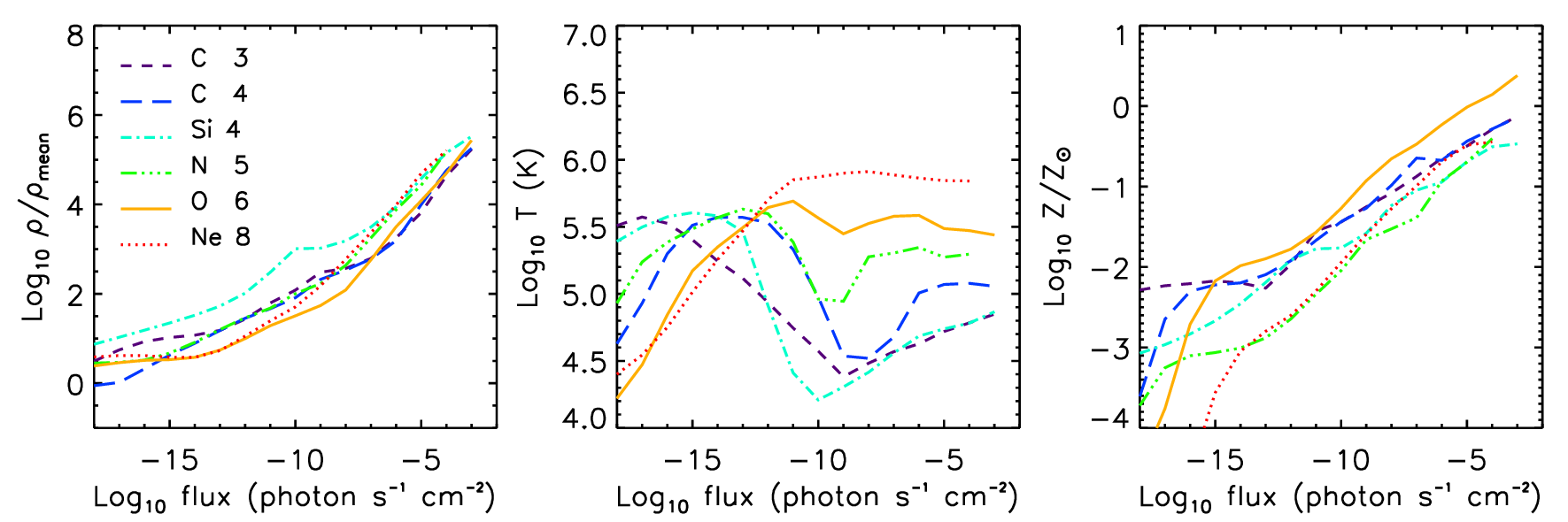}
\caption{Median values of the particle gas density (left panel), temperature (middle panel) and metallicity (right panel) that produce a given UV particle flux at $z=0.25$ for the lines \ciii, \civ, \siiv, \nv, \ovi\ and \neviii. Note that fluxes are given in units of photon s$^{-1}$ cm$^{-2}$, but that previous figures used \phot. For reference, $\rho_{\rm mean}(z=0.25)$ corresponds to $n_{\rm H}\approx 4\times 10^{-7}\,\cm^{-3}$ and our star formation threshold corresponds to $\log_{10} \rho/\rho_{\rm mean} \approx 5.4$. Emission from higher density gas (i.e.\ the ISM) was ignored. The median density and metallicity increase with luminosity, while the median temperature flattens out at the temperature for which the emissivity of each line peaks.}
\label{correl}
\end{figure*}

The distributions in Fig. \ref{uvcont} show what gas produces most of the emission in each line. However, a large fraction of the energy is emitted at low fluxes that are impossible to observe in the foreseeable future. It is therefore interesting to investigate the typical properties of the gas that produces fluxes that might be within reach of future observations.

We do this in Fig. \ref{correl}, which shows the median density (left panel), temperature (middle panel) and metallicity (right panel) of the gas as a function of the emitted flux. The scatter in the distributions is large at low fluxes, and decreases with increasing flux.
The flux on the $x$-axis is the flux emitted by individual particles and the density, temperature and metallicity are local quantities which cannot be compared directly with the global gas properties that can be inferred observationally for the ICM and IGM. This is particularly relevant for the gas metallicity, because in the simulations the metal distribution is very inhomogeneous on small scales.

\begin{figure*}
\centering
\includegraphics[height=\textwidth, angle=90]{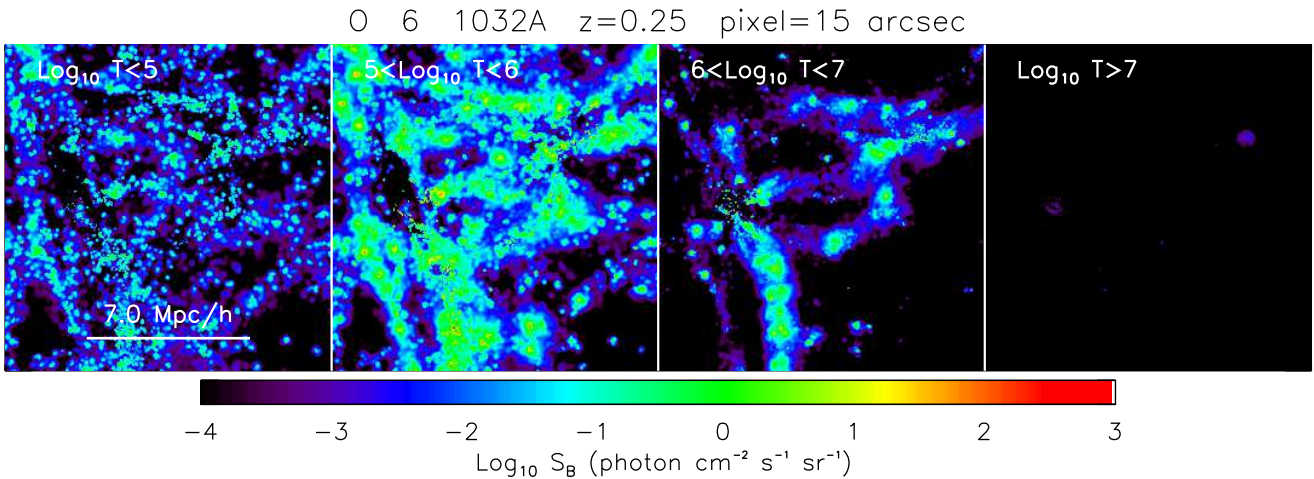}
\caption{As Fig.~\protect\ref{uv_maps}, but showing maps of \ovi\ emission from gas in four different temperature ranges. From left to right: $T<10^5\,\K$, $10^5\,\K$ $<T<10^6\,\K$, $10^6\,\K$ $<T<10^7\,\K$ and $T>10^7\,\K$. The highest \ovi\ fluxes are produced by gas with $10^5$ K $<T<10^6$ K.}
\label{tempcut}
\end{figure*}

As observed in Paper I, the median density and metallicity of the gas increase steadily with flux, as would be expected from the $Z\rho^2$ dependence of the emissivity. On the other hand, the median gas temperature varies strongly for the lowest flux, then flattens out at the highest fluxes with $F\ga 10^{-10}$ photon s\1\ cm\2. This confirms the finding of Fig. \ref{uvcont} that most of the emission comes from gas with temperatures close to the peak temperature of the emissivity curve. Indeed, the highest fluxes are produced by gas in a narrow temperature range centred on this value. As such, the detection of emission lines gives stringent constraints on the gas temperature and provides information on both the density and metallicity of the gas. \ciii\ and \siiv\ lines are the only exceptions to this rule, and show a slight increase in the median temperature at the highest fluxes, while never fully reaching a plateau. The detection of multiple lines may further help to break the degeneracy between density and metallicity.

We have verified that the trends with density and temperature persist if metallicity is assumed to be constant.

\subsection{Density and temperature cuts}
\label{cuts}

\begin{figure*}
\centering
\includegraphics[height=\textwidth, angle=90]{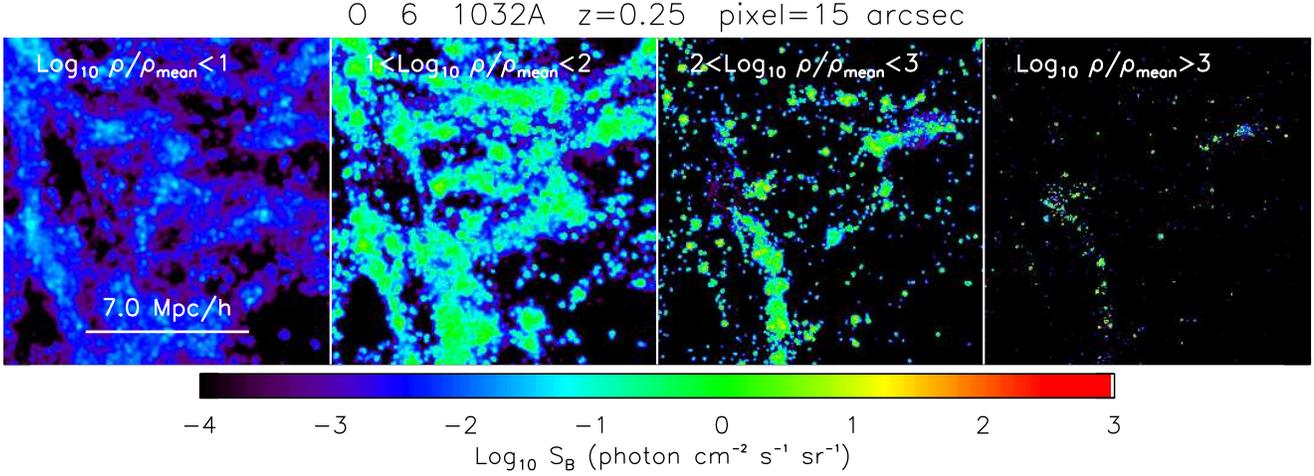}
\caption{As Fig.~\protect\ref{uv_maps}, but showing maps of \ovi\ emission for gas in four different density ranges, as indicated at the top of each panel. The highest fluxes are associated with the densest gas.}
\label{dencut}
\end{figure*}

In this Section we create \ovi\ emission maps using only gas with temperatures and densities in a given interval. This allows us to probe the dependence of the emission on temperature and density and to visualise how the emitting gas is distributed in space.

\begin{figure}
\centering
\includegraphics[width=8.4cm]{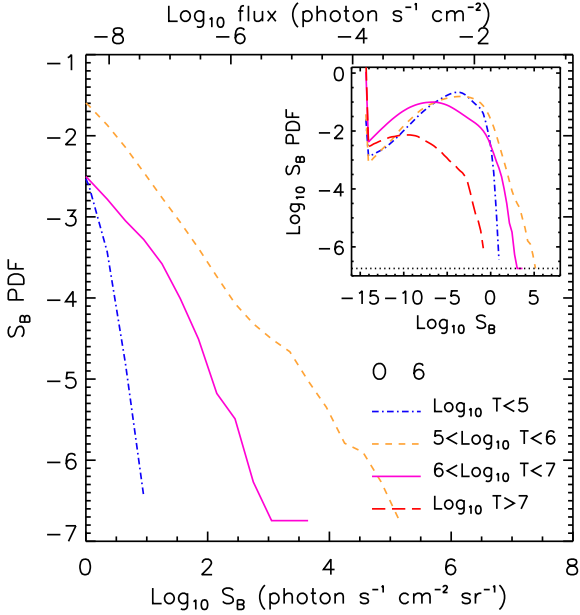}
\caption{As Fig. \ref{uv_pdf}, but showing the PDF of the \ovi\ surface brightness from gas in different temperature intervals.
The highest \ovi\ flux is produced by gas with $10^5$ K $<T<10^6$ K, while the emission is weakest in the highest temperature range. The detailed flux distributions vs. temperature vary line by line, and strongly depend on the peak of the emissivity curve of each line, with the strongest fluxes associated with the temperature range where the emissivity peaks.}
\label{temp_pdf}
\end{figure}

\begin{figure}
\centering
\includegraphics[width=8.4cm]{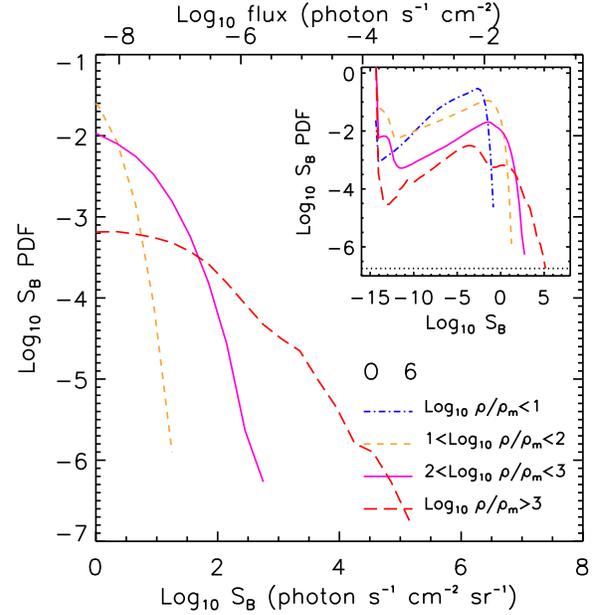}
\caption{As Fig.~\protect\ref{uv_pdf}, but showing the PDF of the \ovi\ surface brightness from gas in different density intervals. The highest fluxes are associated with the densest gas. In particular, the maximum surface brightness predicted for gas outside the virial radii of haloes (i.e.\ $\rho / \rho_{\rm mean} \lesssim 200$) is at least a few orders of magnitude weaker than the maximum surface brightness from denser regions.}
\label{dense_pdf}
\end{figure}

Fig. \ref{tempcut} shows \ovi\ emission maps made using only gas particles that have temperatures in a specific range. We consider the four temperature intervals $T<10^5$ K, $10^5$ K $<T<10^6$ K, $10^6$ K $<T<10^7$ K and $T>10^7$ K. Similarly, in Fig. \ref{dencut} we consider four different density intervals, with $\textrm{Log}_{10} \rho / \rho_{\rm mean}$ varying by one dex in each bin. The surface brightness PDFs for the same cuts are shown in Fig. \ref{temp_pdf} for the temperature cuts and in Fig. \ref{dense_pdf} for the density cuts.

Figs. \ref{tempcut}--\ref{dense_pdf} demonstrate that the strongest \ovi\ emission comes from gas in dense ($\textrm{Log}_{10} \rho / \rho_{\rm mean} > 2$) and moderately hot ($10^5$ K $<T<10^6$ K) regions.
Gas with $10^6$ K $<T<10^7$ K contributes a significant fraction of the bright emission, while cooler gas with $T<10^5$ K contributes only a minor fraction of the total emitted energy. The \ovi\ emission of hot gas with with $T>10^7$ K is negligible. Comparing Fig. \ref{tempcut} with Fig. \ref{dencut} shows that the warm gas emitting in \ovi\ can be identified with the circum-galactic gas. Observing the diffuse filamentary gas ($10^5$ K $<T<10^6$ K; $1<\rho/\rho_{\rm mean} < 2$) would require detection of emission with $S_{\rm B}\sim 1$ \phot.

\section{Dependence on the physical models}
\label{em_model}

\begin{figure*}
\centering
\includegraphics[width=0.8\textwidth]{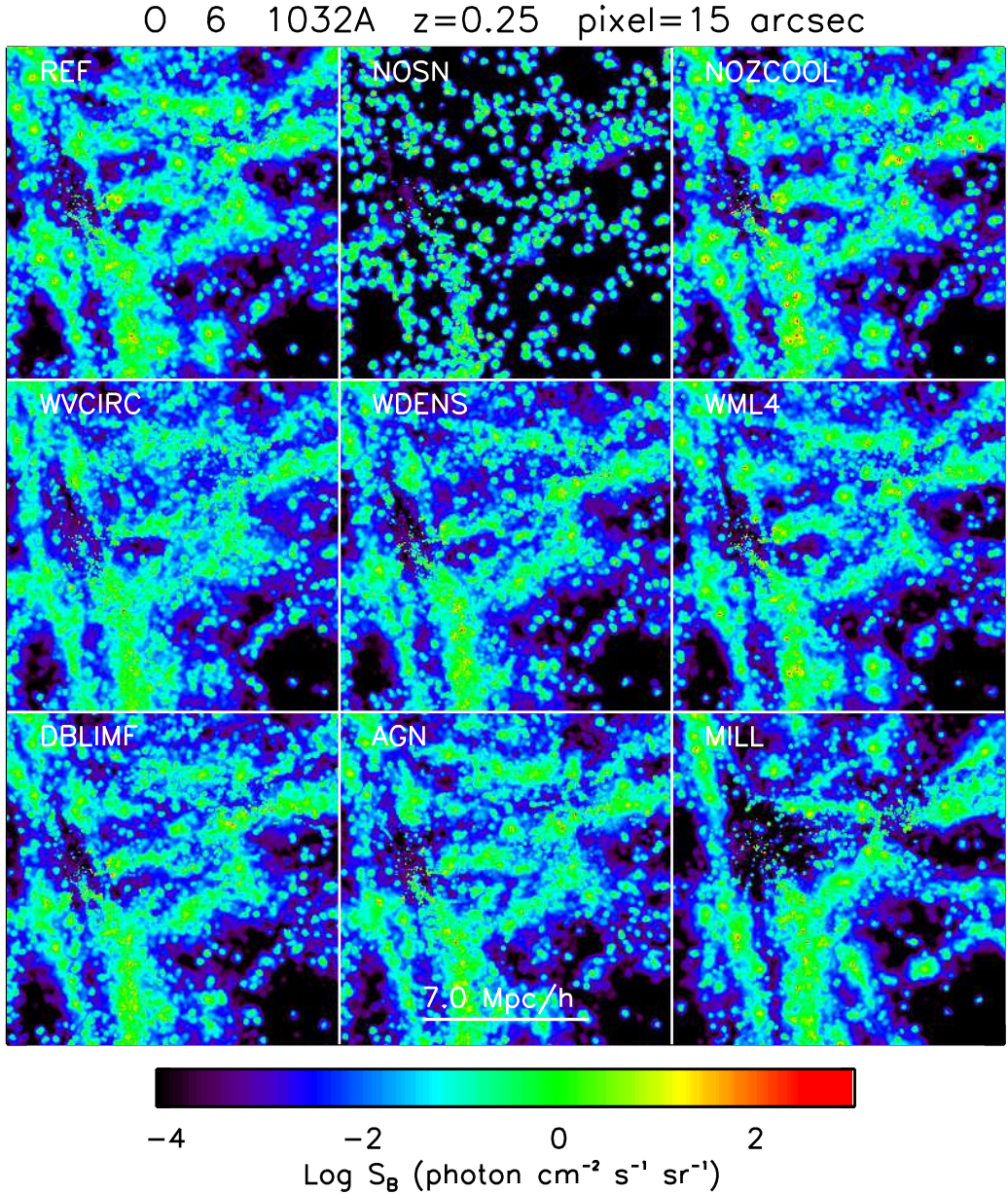}
\caption{As Fig.~\protect\ref{uv_maps}, but showing \ovi\ emission maps for different simulations, as indicated in the top left corner of each panel.}
\label{sims}
\end{figure*}

\begin{figure*}
\centering
\includegraphics[width=0.8\textwidth]{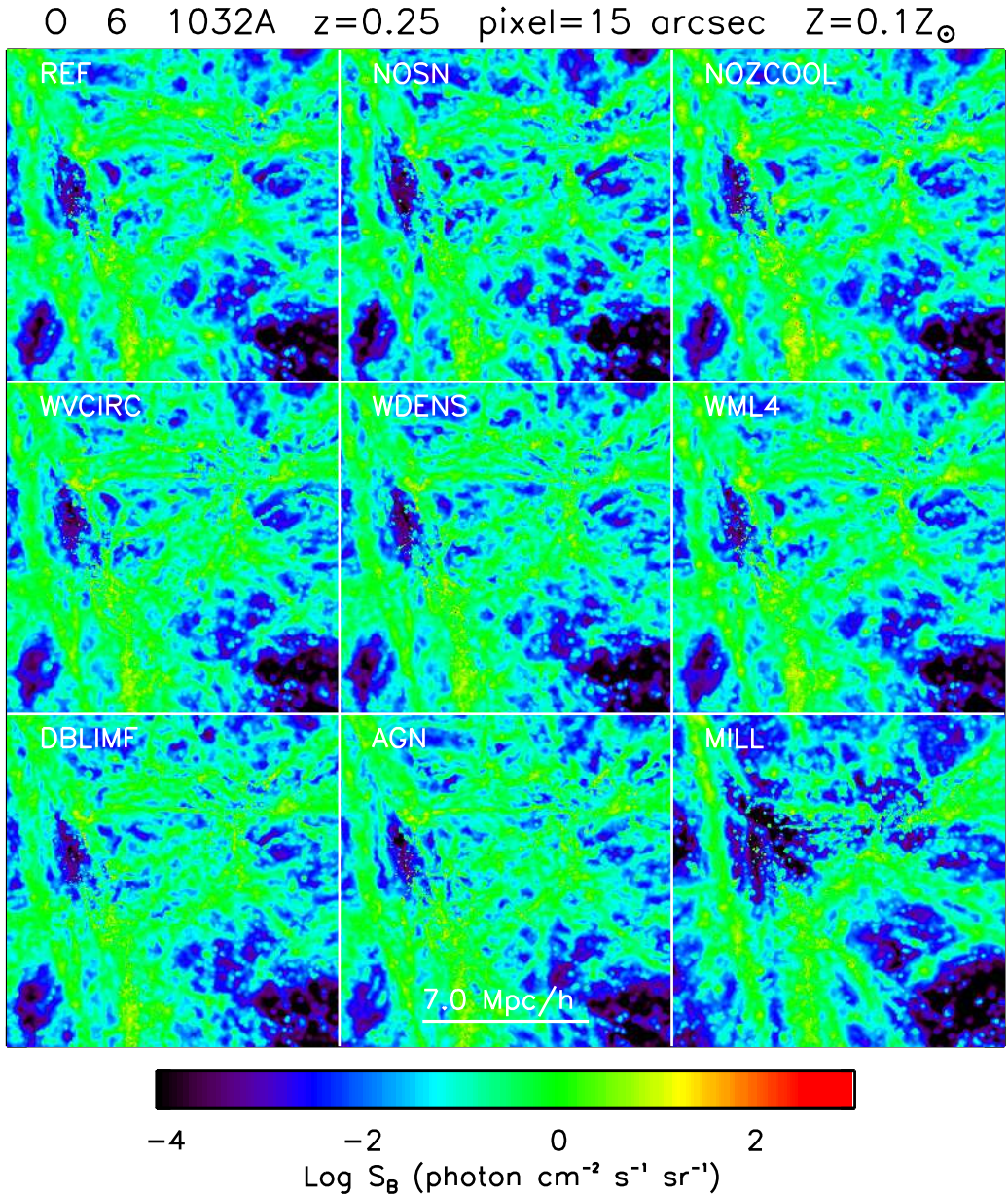}
\caption{Same as Fig. \ref{sims}, but for constant gas metallicity $Z=0.1 Z_{\sun}$.}
\label{simszconst}
\end{figure*}

\begin{figure*}
\centering
\includegraphics[width=0.49\textwidth]{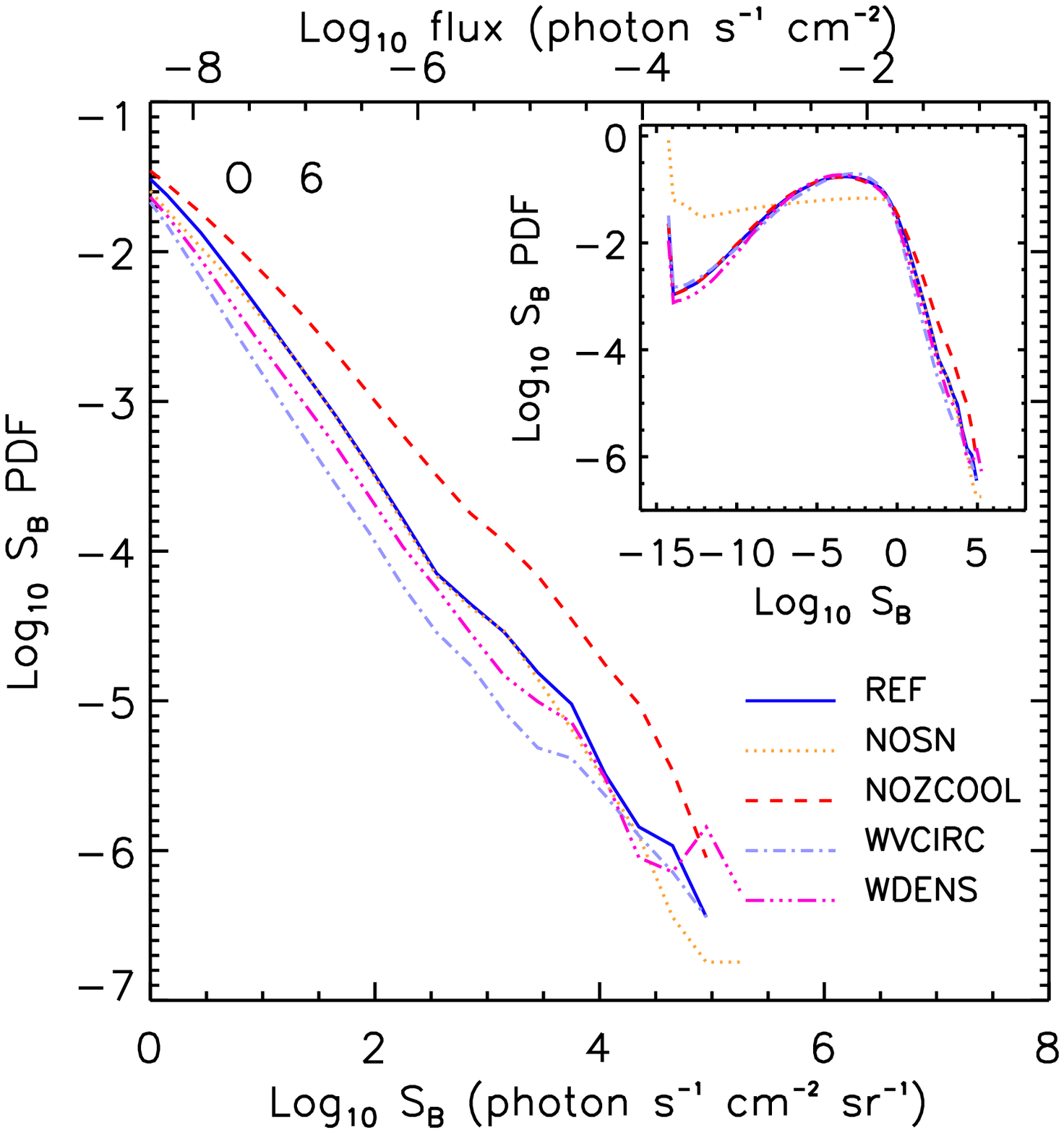}
\includegraphics[width=0.49\textwidth]{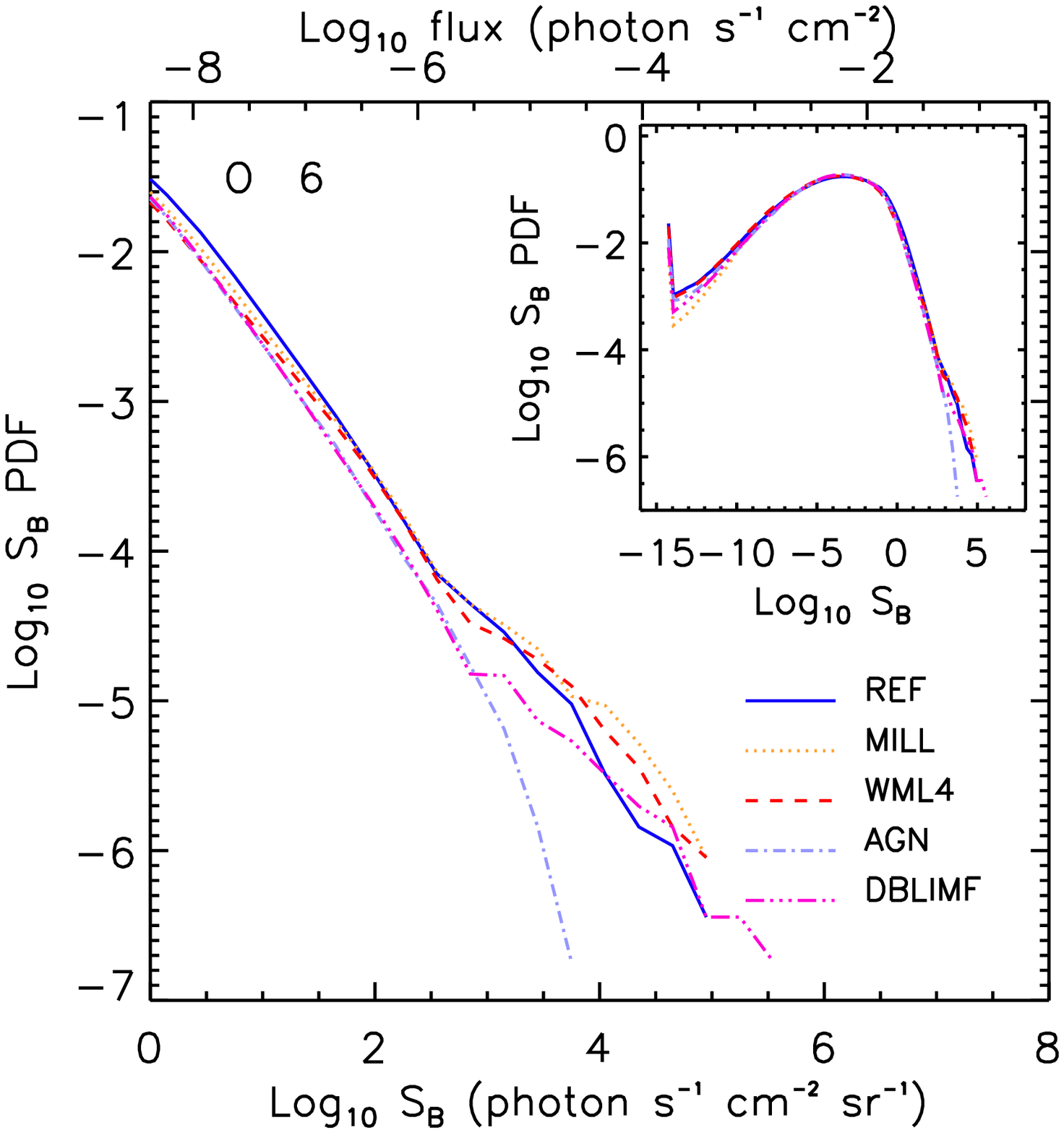}
\caption{As Fig.~\protect\ref{uv_pdf}, but showing the surface brightness PDFs of \ovi\ emission for simulations with varying physical prescriptions. The simulations are the same as in Fig.~\ref{sims}. The results are surprisingly robust to changes in the physical model. However, neglecting metal-line cooling results in an overestimate of the flux at the bright end. Ignoring supernova-driven winds has little effect for the brightest pixels, but otherwise shifts the PDF to lower fluxes. AGN feedback significantly reduces the maximum flux.}
\label{sims_pdf}
\end{figure*}

\begin{figure*}
\centering
\includegraphics[width=0.49\textwidth]{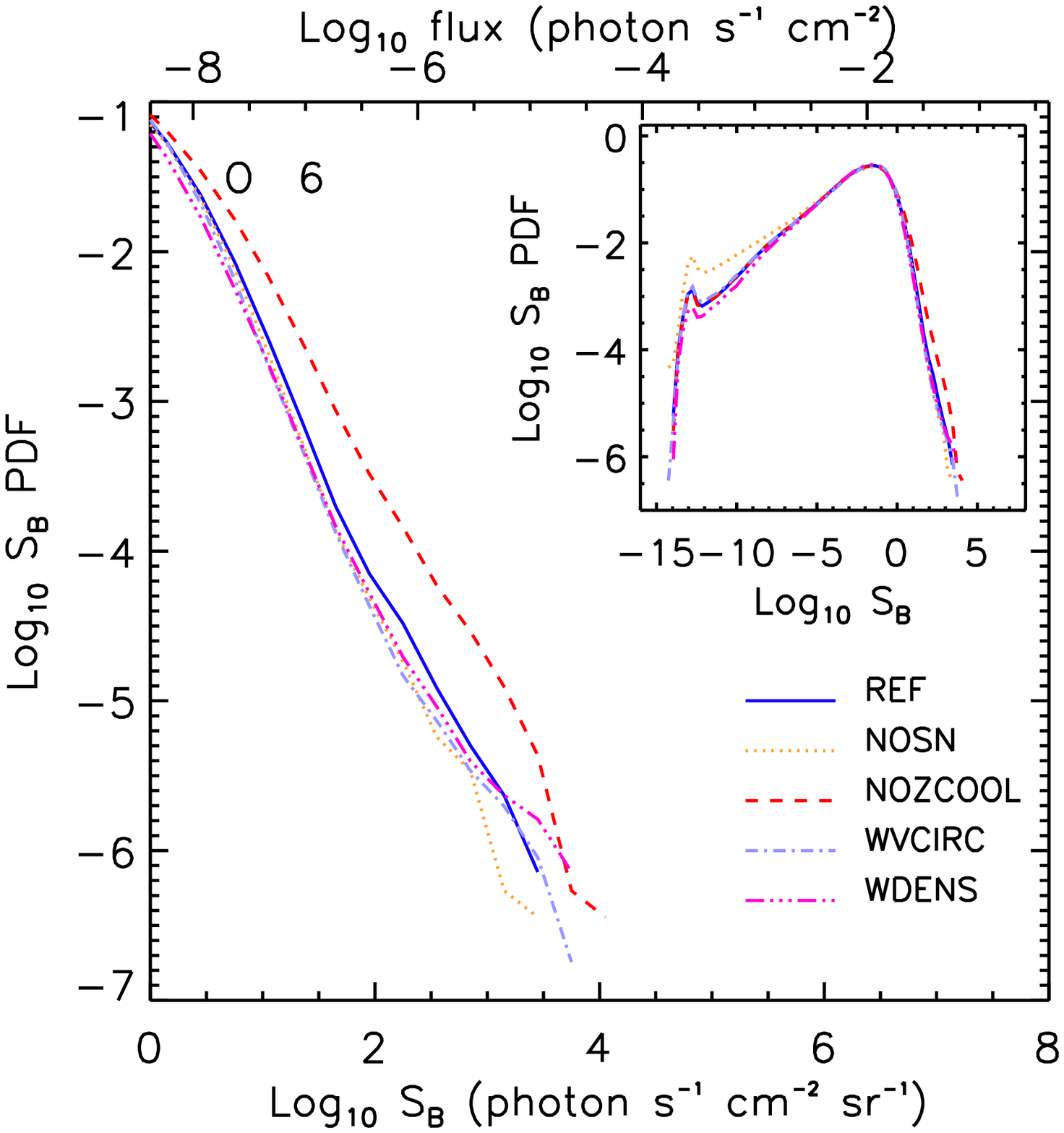}
\includegraphics[width=0.49\textwidth]{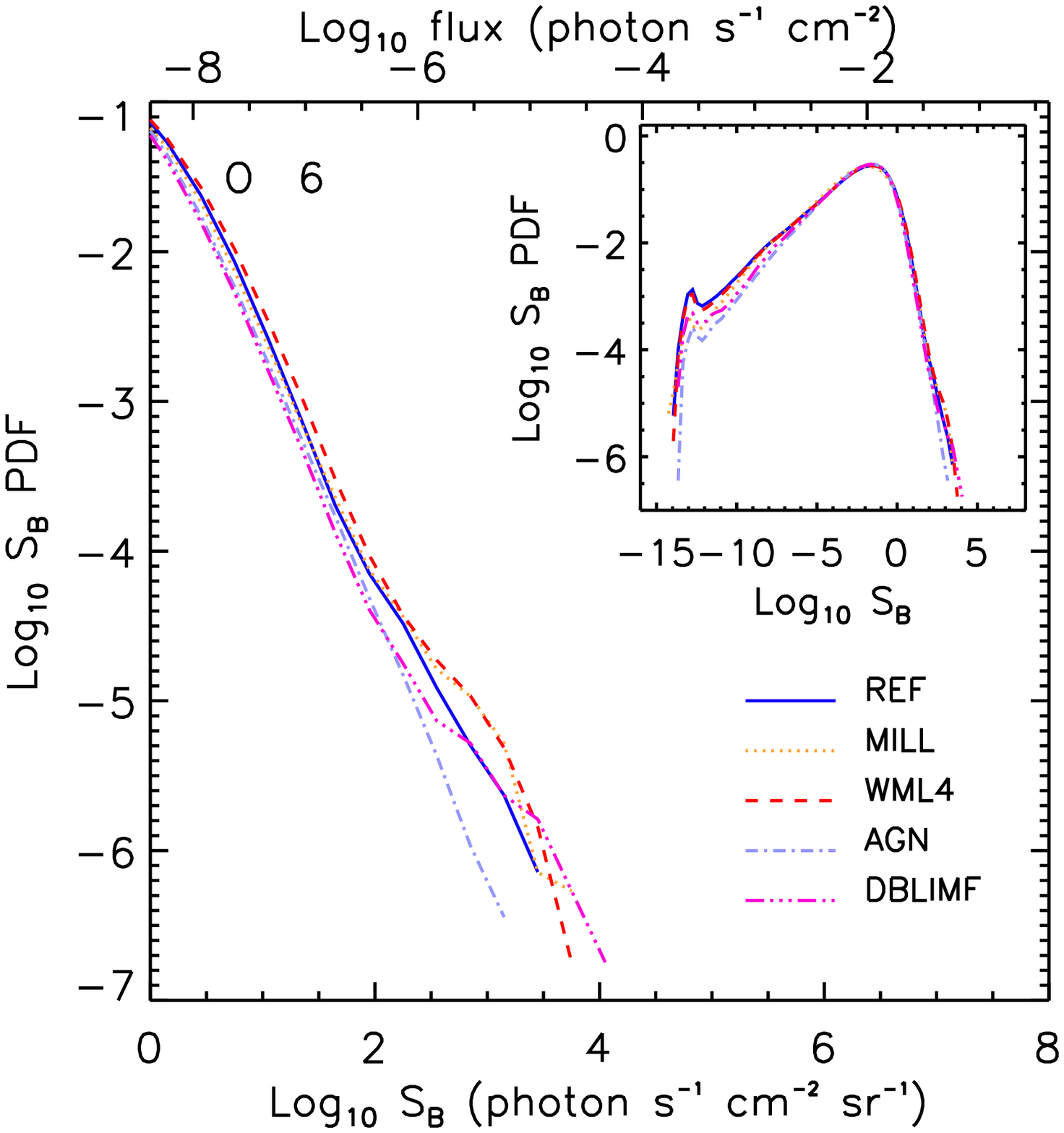}
\caption{As Fig.~\protect\ref{sims_pdf}, but for simulations with varying physical prescriptions, under the assumption of a constant gas metallicity of $Z=0.1 Z_{\sun}$. The simulations shown are the same ones as in Fig.~\ref{sims}. The assumption of constant metallicity boosts the flux from the low-density, cool regions of the universe, where the largest contribution to the line emissivity comes from photo-ionisation by the UV background (compare the inset to that of Fig.~\ref{sims_pdf}), but it reduces the maximum fluxes. The differences between the models are qualitatively the same as when the predicted metal distributions are used (compare with Fig.~\ref{sims_pdf}).}
\label{zconst_sims_pdf}
\end{figure*}

The principal feature of the \owls\ project is the availability of a large number of runs with varying physical prescriptions. In this Section we investigate how a number of physical implementations affect the predictions for the WHIM emission. We present results for \ovi\ $\lambda 1032$ \AA\ emission, but note that the same qualitative conclusions hold for the other lines considered in this study. We focus on models with varying feedback schemes and cooling recipes, which are the physical implementations that produce the largest variations for the diffuse emission. Physical prescriptions are changed one at a time, in order to isolate their impact on the results.
The physical models considered here are the following (we refer the reader to \citet{schaye2010} and to the works cited below for a more detailed description of the specific physical implementations):

\begin{enumerate}
\item \default: Our reference model, as described in Section~\ref{owls}. All results presented in the preceding sections were obtained from this simulation. This model uses the star formation prescription  of \citet{schaye2008}, the supernova-driven wind model of \citet{vecchia2008} with wind mass loading $\eta = 2$ (as defined in Section \ref{owls}) and wind initial velocity $v_{\rm w}=600$~\kms, the metal-dependent cooling rates of \citet{wiersma2009a}, the chemical evolution model of \citet{wiersma2009b} and the WMAP3 cosmology. AGN feedback is not included.

\item \nosn: As \default, but without supernova
  feedback. The metals produced by stars are only transported by gas
  mixing and no energy is transferred to the IGM and ISM when SNe
  explode. This model is clearly unrealistic, it for example predicts far more star formation than obseved \citep{schaye2010}. We include it, however, because it can shed light on the physics that determines the metal-line emission. 

\item \zcool: As \default, but with radiative cooling rates calculated assuming primordial abundances. The calculation of the metal line emission in this model is not self-consistent, since the contribution of metals to the cooling rate is not included. The neglect of the contributions of metal lines implies slower gas cooling than in the \default\ model. We include this model to investigate the role of metal-line cooling and because previous work often assumed primordial abundances \citep[e.g.][see Section~\ref{furla}]{furlanetto2004}.

\item \agn: As \default, but with the addition of AGN feedback as described in \citet{booth2009}. We note that \citet{mccarthy2010} have shown that this model (but not model \default) reproduces the observed density, temperature, entropy and metallicity profiles of groups of galaxies, as well as their star fractions. Model \agn\ may therefore, for our purposes, be the most realistic model of the ones considered here.

\item \wml: As \default, but with a wind mass loading twice as high, that is $\eta = 4$. The total energy injected in winds is also twice that in \default.

\item \wdens: As \default, but the wind properties scale with the local
  gas density as:
\begin{eqnarray}
v_{\rm w} &=& v_{\textrm{w}0} \left(\rho / \rho_{\rm crit} \right)^{1/6}\\
\eta &=& \eta_0 \left( \rho / \rho_{\rm crit}\right)^{-2/6}
\end{eqnarray}
where $\rho_{\rm crit}$ is the density threshold for star
formation, $v_{\textrm{w}0}= 600$~\kms\ and
$\eta_0 = 2$. This implies that at $\rho_{\rm crit}$ the gas is kicked
with the same velocity and mass-loading as in the \default\ model.
This particular scaling of wind velocity with density was chosen so
that the wind velocity scales with the local sound speed (recall that
we impose the equation of state $P\propto \rho^{4/3}$ onto the ISM) and the
wind mass-loading is scaled such that the total wind energy is
independent of the gas density. As a consequence, the wind energy per unit stellar mass is equal to that in the \default\ model.

\item \wmom: As \default, but with ``momentum--driven'' galactic winds
  following the prescription of \citet{oppenheimer2008}, with the
  difference that winds are ``local'' to the star formation event and
  are fully hydrodynamically coupled as in the \default\ model (see
  \citealt{vecchia2008} for a discussion of the importance of these
  effects).  The 
  wind initial velocity and the wind mass loading factor are defined
  as a function of the galaxy velocity dispersion $\sigma$ as $v_{\rm
    w} = 5\sigma$ and $\eta = v_{\textrm{w}0}/\sigma$ respectively,
  with $v_{\textrm{w}0}=150$~\kms. The velocity dispersion $\sigma =
  \sqrt{2}v_{\rm circ}$ is estimated using an on-the-fly
  friends-of-friends halo finder. This model was motivated by the idea
  that galactic winds may be driven by radiation pressure on dust grains
  rather than SNe. The energy injected into the wind
  becomes much higher than for \default\ for halo masses exceeding $10^{11}
  - 10^{12}\,{\rm M}_\odot$.

\item \mill: As \default, but using the WMAP year 1
  cosmology \citep{spergel2003} as in the Millennium simulation
  \citep{springel2005} and a wind mass loading factor $\eta =
  4$. Results from this model must be compared to those of the
  \wml\ model to isolate the effect of cosmology.

\item \dblimf: As \default, but using a top-heavy IMF at high gas
  pressures. This model switches from a Chabrier IMF to a power-law
  IMF $dN/dM\propto M^{-1}$ (as compared to 
  $\propto M^{-2.3}$ for the high-mass tail of the Chabrier IMF) at
  the critical
  pressure $P/k=2.0\times 10^6\,\cm^{-3}\,\K$, which was chosen
  because for this value $\sim 10^{-1}$ of the stellar mass forms at
  higher pressures. The extra SN energy is used to increase the
  initial wind velocity to $1618$~\kms for star particles formed with a
  top-heavy IMF, while keeping the mass loading factor fixed at
  $\eta=2$ (as for the Chabrier IMF, these parameter values
  correspond to 40\% of the available SN energy). This model is called
  \emph{DBLIMFCONTSFV1618} in the \owls\ project (which includes
  more models with top-heavy IMFs in starbursts, see
  \citealt{schaye2010}). 

\end{enumerate}

Fig.~\ref{sims} shows maps of the \ovi\ emission for all runs in a region of 14 \hm\ Mpc on a side, or equivalently 1.12 degrees on the sky. Fig. \ref{simszconst} shows the emission in the same simulations, but assuming a constant gas metallicity of $Z=0.1 Z_{\sun}$.
The corresponding surface brightness PDFs of the \ovi\ emission are presented in Fig. \ref{sims_pdf}, while those for constant metal abundances are shown in Fig. \ref{zconst_sims_pdf}.

Despite the substantial differences in the physical models, we find that most models produce relatively small differences in the results. In particular, the specific implementation of stellar feedback (\wmom, \wdens) barely affects either the surface brightness PDF (Fig.~\ref{sims_pdf}). Even the small differences between feedback models found in Paper I for the \xray\ emission are mostly washed out when considering lower ionisation transitions in the UV band.  The only exceptions that show noticeable differences with the \default\ model are \nosn, \zcool\ and \agn\, and we now discuss each of these in turn. 
 
The spatial distribution of the emission for the \nosn\ model is more clumpy and concentrated around galaxy haloes than for all other models. The flux PDF of the \nosn\ model differs primarily at low fluxes ($< 1$ \phot). This is a consequence of the fact that without SN winds the transport of metals to the IGM is very inefficient. This naturally produces very low IGM metallicities, and therefore low metal line emission away from galaxies. Although the surface brightness distribution predicted by the \nosn\ model differs greatly from that in the runs that include SN feedback (Fig.~\ref{sims}), when the same maps are made assuming constant gas metallicity (Fig.~\ref{simszconst}) most of the differences disappear. This indicates that the primary way in which SN feedback affects the UV emission is through altering the metal distribution and not necessarily by changing the thermodynamic properties of the gas emitting in the UV band.

The \zcool\ run predicts a larger fraction of pixels with high fluxes than all other models. This indicates that the \zcool\ model predicts a larger fraction of gas within the temperature range that produces \ovi\ emission.

The surface brightness PDF for the \agn\ model is remarkably similar to those of other models for all but the highest fluxes ($ >100$ \phot), where it shows a steep cut-off. This indicates that the flux from high density regions is lower than in the other models and is likely to be a consequence of two physical processes.  Firstly, AGN feedback suppresses star formation in haloes that host black holes, which in time would substantially reduce the gas metallicity.  Secondly, as demonstrated by \citet{mccarthy2010} and \citet{duffy2010}, the inclusion of AGN feedback in the simulations significantly decreases the gas fractions in the centres of all massive haloes.  Together, decreases in both the density and metallicity of the gas lead to greatly suppressed fluxes in the highest density regions.

Variations in the cosmology, seen by comparing the \mill\ and \wml4\ models, do not produce significant variations in the PDF, but do affect the spatial distribution of the emission, as can be seen in Fig. \ref{sims}. This is mostly a consequence of the fact that structure formation has progressed further in the \mill\ run due to the higher value of $\sigma_8$ used in this simulation.

If we assume a constant gas metallicity, as in Figs. \ref{simszconst} and \ref{zconst_sims_pdf}, then the results barely vary with varying physical prescriptions, with the exception of the \zcool\ model, whose results reflect those for the case with the simulation metallicities.

In summary, except for metal-line cooling and AGN feedback, the physical prescriptions we have tested here only marginally affect the properties of the gas reservoir that produces UV emission, which is therefore insensitive to variations in the SN feedback scheme and stellar IMF (as in the \dblimf\ model). The results presented in this work are therefore robust against uncertainties in the subgrid models discussed in this section, although AGN feedback may significantly suppress the peak fluxes.

\section{Comparison with previous work}
\label{furla}

Our predictions for \civ\ and \ovi\ emission can be compared directly to those of Furlanetto et al. (2004, F04 hereafter, Figs.~7--10), who use the ``G5'' simulations of \citet{springel2003}. In particular, Fig.~10 of F04 uses the same box size and redshift as our Figs.~\ref{uv_pdf} and \ref{sims_pdf}. F04 use a slice thickness of $\Delta z = 0.01$, whereas $\Delta z \approx 0.0074$ for us. Since the bright-end of the surface brightness PDF is proportional to the slice thickness (see appendix~\ref{thick}), we have reduced their PDFs by about 0.13~dex for a fair comparison. They use a pixel size of 20 proper kpc$/h$, which is about half the size of our pixels. However, Fig.~\ref{angle_figure} shows that this difference has little effect on the results. The simulation of F04 uses about four times less particles, but our Fig.~\ref{number_figure} indicates that this is sufficient. They use a WMAP1 cosmology instead of WMAP3, but Fig.~\ref{sims_pdf} shows that this also does not make a significant difference. The simulation used by F04 used different prescriptions for chemodynamics (they did not track individual elements and they did not follow the delayed release of heavy elements by stars), star formation and SN feedback and, like our reference model, did not include AGN feedback. The most important difference is, however, likely the fact that F04 did not include metal-line cooling. 

Comparing F04's prediction for \ovi\ to that for our \default\ model, we find that bright regions are somewhat more common in their simulation. However, since they did not include metal-line cooling, it is more appropriate to compare with our model \zcool. Reassuringly, in that case the bright ends of the PDF agree nearly exactly. The same, is not true, however, for \civ. While F04 predict the maximum fluxes to be weaker for \civ\ than for \ovi, we find the opposite. Since our predictions agree for \ovi\ (if we ignore metal-line cooling, which is not self-consistent), this means we predict substantially more \civ\ emission. This could be because F04 did not include metal enrichment by SN~Ia and AGB stars. While this is not that important for oxygen, which is released mostly by massive stars, intermediate mass stars are important for carbon.

\section{Can the WHIM be detected in the UV?}
\label{uv_em}

In Section \ref{cuts} we showed that UV emission lines such as \civ, \nv, \ovi\ and \neviii\ are good tracers of the coolest fraction of the WHIM and tend to be strongest in the haloes of galaxies and outside the central cores of groups. In this Section, we discuss their observability by current and future instruments, in particular by \hstcos\  \citep{froning2008}, \fireball\  \citep{tuttle2008} and \atlast\  \citep{postman2008}.
The requirements for UV telescopes to detect metal-line emission lines from the WHIM are the same as those for \xray\ instruments: a large field of view (FoV) plus high spatial and angular resolution.

\hstcos, which was successfully mounted on the Hubble Space Telescope, has been designed to detect WHIM absorption features, rather than diffuse emission. Its extremely small FoV is optimal to detect light from faint UV point sources, but its spectral resolution declines steeply for extended sources. The flux limit of \hstcos, of order 1000 \phot, is too high for allowing the detection of WHIM emission in filaments or outside groups at $z=0.25$, which is of the order of $0.01-1$ \phot. However, it might just be sufficient to detect metal-line emission from the densest gas in the haloes of galaxies.

The \fireball\ balloon experiment \citep{tuttle2008} has been successfully launched in June 2009, with the specific aim of detecting diffuse emission from \civ\ at $z\sim 0.3$ and \ovi\ at $z\sim 1$. \fireball\ is a pathfinder integral field unit with a relatively large field of view of $2.5'\times 2.5'$ and an angular resolution of 10". The spectrograph is fed by about 400 fibres sensitive in the narrow UV band 1950\AA -- 2200\AA, and can reach a flux limit of $\sim 2000$ \phot. Such a flux limit is substantially lower than the maximum value of $\sim 10^5$ \phot\ we predict for \civ\ emission at $z\sim 0.25$ (see Fig. \ref{uv_pdf}). For \ovi\ we predict a maximum surface brightness at $z\sim 1$ of $\sim 3000$ \phot\  (see Fig. \ref{redshift_pdf}), which is close to the detection threshold.
However, AGN feedback may decrease these maximum fluxes by an order of magnitude (Fig. \ref{sims_pdf}).
As such, we predict that \fireball\ should be capable of detecting a significant number of photons associated with diffuse \civ\ emission from high density regions. Detection of \ovi\ emission from high density regions might also be feasible, but this represents a more challenging prospect at the very limits of the instrument capabilities.
\ciii\ is predicted to be even stronger than \civ, but this line does not trace the WHIM and, being a singlet, it would be difficult to separate from contaminating \hi\ Ly series lines (e.g.\ \citealt{furlanetto2005}).
We are looking forward to seeing the first results. A positive detection of diffuse emission would certainly be a powerful argument in favour of future larger UV telescopes.

Ideally, next generation UV missions targeted to detect diffuse emission from the WHIM should aim to achieve flux limits of the order of 1--10 \phot\ within a FoV with size of at least an arcminute. Such a low flux limit is necessary to detect emission lines from truly diffuse gas outside the haloes of galaxies. The proposed specifications for the \atlast\ mission \citep{postman2008} may be ideally suited for the detection of WHIM emission from filaments.

Finally, we note that we have not considered emission from hydrogen Lyman series lines, because they are thought to trace cooler gas than the WHIM. However, the intensity of \lya\ emission may well be higher than that of the metal lines and Lyman series lines are therefore a potential contaminant for other rest-frame UV lines \citep[e.g.][]{furlanetto2003,furlanetto2005}. Although doublets can be unambiguously identified with sufficient spectral resolution, singlets like \ciii\ ($\lambda$977~\AA) would be difficult to distinguish from contaminating hydrogen lines unless other lines corresponding to the same redshift are detected.

\section{Conclusions}
\label{summary}

UV metal lines constitute one of the main cooling channels of warm gas. These UV lines provide an attractive route towards detecting the cooler fraction of the missing baryons in the low redshift Universe, which simulations predict to reside in a warm-hot diffuse phase with $10^5$ K $<T<10^7$ K (the WHIM).

In this work, we used a subset of cosmological simulations from the
OverWhelmingly Large Simulations (\owls) project \citep{schaye2010} to
study the nature and detectability of rest-frame UV metal-line emission from diffuse gas ($n_{\rm H} < 10^{-1}\,{\rm cm}^{-3}$). The \owls\ runs
are among the largest dissipative simulations ever performed, and use new
prescriptions for a range of baryonic processes, such as radiative cooling, star
formation, and feedback from SNe and AGN. Of particular relevance is that the \owls\ runs are fully chemodynamical and track the release of 11 different
elements by massive stars, SNe of types II and Ia, and asymptotic giant branch stars. In addition, radiative cooling is implemented element-by-element in the presence of an evolving UV/\xray\ background. 

Our conclusions can be summarised as follows:

\begin{enumerate}

\item In the UV band the strongest metal lines are \ciii\  (977~\AA, $T\sim 10^{5}\,$K), \civ\  (1548,1551~\AA, $T\sim 10^5\,$K) and \ovi\  (1032,1038~\AA, $T\sim 10^{5.5}\,$K). \ciii\ is the strongest line and mostly traces circum-galactic gas. \siiv\ (1393,1403~\AA) traces a fraction of the high density gas traced by \ciii\ emission, within a smaller range around the peak temperature $T\sim 10^{5}\,$K. The maximum \civ\ emission is almost an order of magnitude weaker than the peak \ciii\ emission, and traces gas that has a similar spatial distribution as the gas that produces \ciii\ emission, but is slightly warmer. \ovi\ emission arises from warmer and more diffuse gas than \civ\ emission. As a consequence, \ovi\ emission is weaker than \civ\ close to galaxies, but stronger outside haloes.
The maximum surface brightness for \nv\ (1239,1243~\AA, $T\sim 10^{5.5}\,$K) and \neviii\ (770,780~\AA, $T\sim 10^6\,$K) is weaker than that of \ovi\ by up to an order of magnitude. The properties of \nv\ emission are intermediate between those of \civ\ and \ovi. The \neviii\ line is qualitatively similar to \ovi\ and traces warm-hot gas with $T\sim 10^6\,$K.

\item Emission lines provide strong constraints on the temperature of the emitting gas, because the bright emission traces gas with temperatures close to the peak of the emissivity curve. Photo-ionisation by the UV background is unimportant for the gas that dominates the emission from the intergalactic medium. Since this temperature increases with atomic number, \civ\ lines trace cooler gas than \ovi\ emission, which in turn comes from cooler gas than \neviii\ emission.

\item UV emission lines are likely to be detectable only in relatively high density regions. The detection of gas with density $\rho < 10^3 \rho_{\rm mean}$ requires $S_{\rm B}\la 10^2$ \phot, while gas with $\rho < 10^2 \rho_{\rm mean}$ requires $S_{\rm B}\la 10$ \phot. The gas that would be detected at these surface brightness levels tends to be relatively metal-rich ($Z \ga 10^{-1}\,Z_\odot$).
The detection of low density structures such as filaments ($\rho \sim 10 \rho_{\rm mean}$) is thus extremely challenging and might be best achieved through absorption techniques. However, the \fireball\ balloon experiment \citep{tuttle2008} has a good chance to detect diffuse \civ\ emission at $z\sim 0.3$. We will present comparisons with absorption line data elsewhere

\item By comparing results from different \owls\ models, we have investigated the dependence of the results on a number of physical mechanisms.
The variations in the predicted fluxes for different simulations are within a factor of ten. Varying the intensity of feedback or the stellar IMF affects the simulated fluxes by at most a factor of a few. Assuming primordial abundances for the calculation of the cooling rates (which is not self-consistent, but often done in previous work) boosts the flux in the brightest pixels by about an order of magnitude. Simulations that include AGN feedback predict maximum fluxes about an order of magnitude lower than the \default\ model in massive haloes, mostly because of the lower gas metallicity that results from the suppression of star formation. However, AGN feedback has little effect on the surface brightness PDF for values more than an order of magnitude below the maximum value (i.e.\ for $S_{\rm B}\la 10^2$ \phot).

\item The quadratic dependence of
the emissivity on density implies that emission is a highly biased
tracer of the IGM mass and metals. While emission lines are thus not an ideal tool to close the baryon budget, they are excellent tracers of the gas flowing in and out of galaxies. Rest-frame UV (and soft X-ray) lines are sensitive to the gas temperatures that are relevant for galactic winds, cooling flows and cold-mode accretion. Thus, they have the potential to open up a new window onto some of the most poorly understood aspects of the formation and evolution of galaxies. 

\end{enumerate}

By comparing the results presented in this work with those of Paper I, we find that, on average, UV emission lines have a surface brightness two to three orders of magnitude higher than soft \xray\ lines. This can be seen, for example, when comparing \ovi\ and \oviii\ emission. The maximum surface brightness predicted for \ovi, assuming an angular resolution of 15", is about $10^5$ \phot, while for \oviii\ it is $\sim 10^3$ \phot, or two orders of magnitude lower. It has to be noted, however, that the relative strength of the emission lines varies strongly with the gas temperature. As a consequence, \xray\ and UV lines are well suited to probe gas with different properties that may not physically occupy the same regions of space.

Effectively, UV and soft \xray\ metal lines together have the potential to yield a huge amount of information about the WHIM and the circum-galactic medium. While \xray\ lines provide clues about the hottest component with $T\ga 10^6$ K, UV lines trace the cooler fraction of the gas with $T\la 10^6$ K including WHIM, galactic winds, cooling flows and cold mode accretion. Ultimately, the multi-wavelength approach may constrain the density, temperature, metallicity and ionisation state of gas, highlighting the fully 3-dimensional structure of the gas distribution. This would shed new light on a wealth of processes, from the feedback mechanisms responsible for the metal enrichment of the IGM, to the chemical evolution of galaxies, and to the cosmic flows that provide fresh fuel for star formation and black hole accretion.

\section*{Acknowledgments}

We would like to thank Jelle Kaastra, Stephan Frank, Xavier Prochaska, Volker Springel and all other members of the \owls\ Team for useful discussions and help during the development of this project. 
The simulations presented here were run on Stella, the LOFAR
BlueGene/L system in Groningen, on the Cosmology Machine at the
Institute for Computational Cosmology in Durham as part of the Virgo
Consortium research programme, and on Darwin in Cambridge. This work was sponsored by the Dutch
National Computing Facilities Foundation (NCF) for the use of
supercomputer facilities, with financial support from the Netherlands
Organization for Scientific Research (NWO). SB acknowledges support
by NSF Grants AST-0507117 and AST-0908910.
This work was supported by Marie
Curie Excellence Grant MEXT-CT-2004-014112, by an NWO VIDI grant, and
was partly carried out under the HPC-EUROPA project, with the support
of the European Community - Research Infrastructure Action of the FP7.

\appendix

\section{Convergence tests}
\label{converge}

In this Appendix we investigate how a number of numerical issues affect our results. In particular, in Appendix \ref{number} we vary the mass resolution of the simulations, while in Appendix \ref{box} we keep the mass resolution constant and vary the size of the simulated volume. Finally, in Appendix \ref{thick} we show how varying the slice thickness affects the shape of the surface brightness PDF. All tests use the \default\ physical model and results are shown only for the \ovi\ 1032 \AA\ line.

\subsection{Mass resolution}
\label{number}

\begin{figure}
\centering
\includegraphics[width=8.4cm]{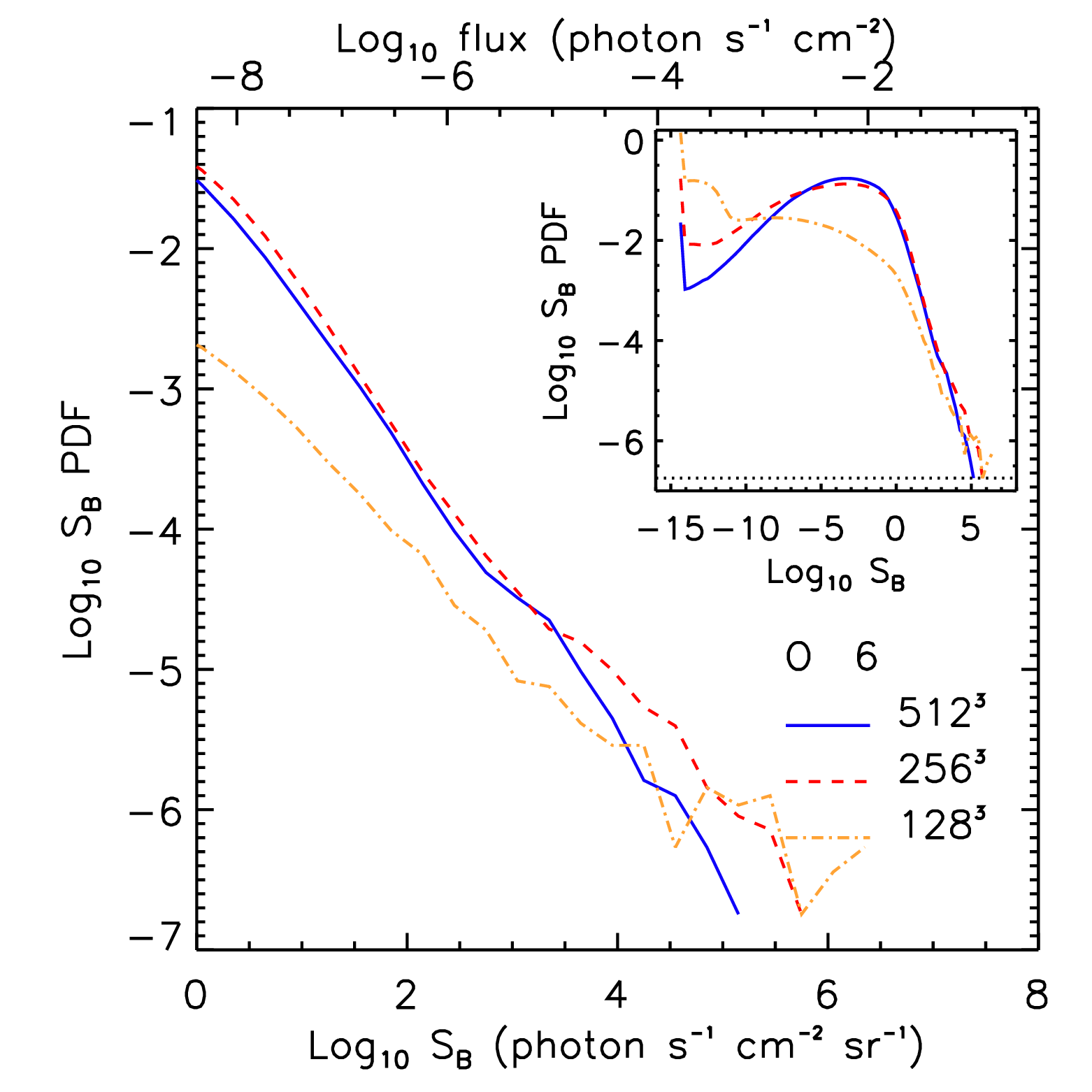}
\caption{As Fig.~\protect\ref{uv_pdf}, but showing the \ovi\ surface brightness PDF as a function of the resolution of the simulation. The particle mass (gravitational softening) increases by a factor of eight (two) when the number of particles decreases by a factor of $2^3$. All runs assume the same physical model, angular resolution and box size. The predictions of the $N=256^3$ and $512^3$ runs are very close for fluxes greater than $10^{-5}\,$\phot, but diverge again for the small number of pixels with fluxes greater than $10^{3.5}\,$\phot.}
\label{number_figure}
\end{figure}

The effect of varying the numerical resolution of the simulations is shown in Fig. \ref{number_figure}, which gives surface brightness PDFs for maps created for three different realisations of the same simulation with particle numbers $N=2\times 128^3$, $2\times 256^3$ and the default number of $2\times 512^3$. The dark matter particle masses are $M_{\rm DM} = 2.6\times 10^{10}$ \hm\  \msun, $3.2\times 10^9$ \hm\  \msun\ and $4.1\times 10^8$ \hm\  \msun, respectively. All runs have a box size of 100 \hm\ Mpc and we assume an angular resolution of 15".

The results for the two high resolution runs with $N=2\times 256^3$ and $2\times 512^3$ particles converge well for intermediate fluxes, but diverge somewhat at the highest and lowest fluxes, with differences within a factor of a few. On the contrary, the lowest resolution run with $N=2\times 128^3$ does not converge anywhere and the overall shape of the PDF is different from those of the other runs, with a lot more power at the lowest fluxes. This is due to the fact that this simulation is not able to properly describe the density and temperature distributions of the gas both on large and on small scales. The star formation history is severely underestimated in the $N=2\times 128^3$ simulation and the metals produced by stars are not efficiently transported into the diffuse IGM. This implies a far lower gas metallicity than in the other runs and explains the very low fluxes observed outside haloes.

\subsection{Box size}
\label{box}

\begin{figure}
\centering
\includegraphics[width=8.4cm]{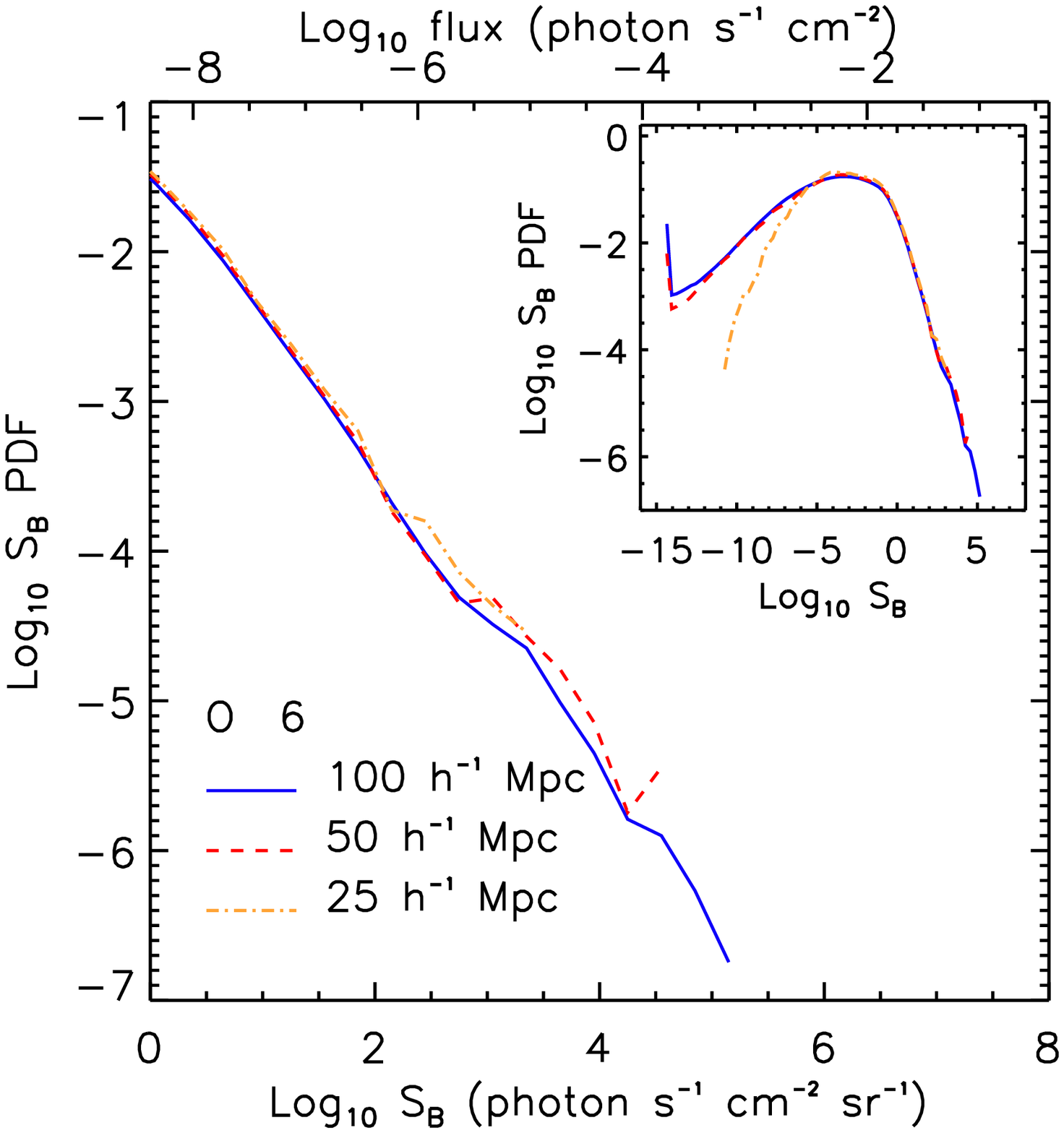}
\caption{As Fig.~\protect\ref{uv_pdf}, but showing the \ovi\ surface brightness PDF for simulations with box sizes of 100, 50 and 25 \hm\ Mpc. All runs assume the same physical model, particle mass and angular resolution. Except for very low fluxes ($<10^{-5}\,$\phot), the results are insensitive to the size of the simulation box. The highest flux values that are sampled increase with box size.}
\label{box_figure}
\end{figure}

The effect of varying the box size of the simulation on the \ovi\ surface brightness PDF is shown in Fig. \ref{box_figure}. The runs assume different initial conditions and allow us to test how the large scale structure affects the flux statistics. The mass resolution in the simulations is kept constant, with a dark matter particle mass $M_{\rm DM} = 4.1\times 10^8$ \hm\  \msun.

We find that the surface brightness PDFs converge, but that the highest fluxes are only sampled in the largest box, in which more extreme objects can form. The surface brightness PDF for the 25 \hm\ Mpc run diverges from the other two at the lowest fluxes.

\subsection{Slice thickness}
\label{thick}
\begin{figure}
\centering
\includegraphics[width=8.4cm]{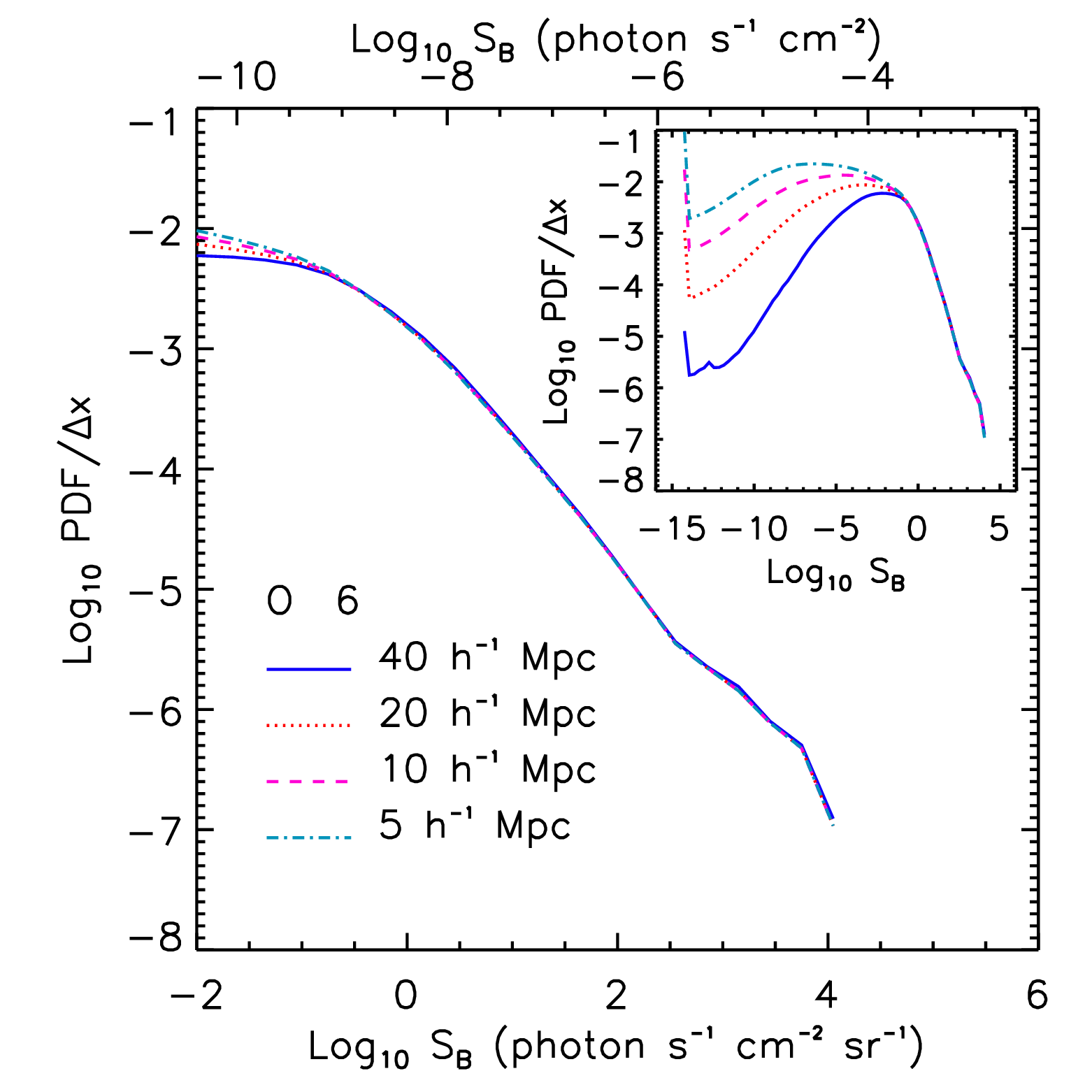}
\caption{The \ovi\ surface brightness PDFs, normalised by the slice thickness $\Delta x$, for maps with varying slice thickness: 40, 20, 10 and 5 \hm\ Mpc. The number of slices used to calculate each PDF is 2, 5, 10 and 20, respectively. The test is for the \default\ model with box size 100 \hm\ Mpc at $z=0.25$. All maps assume the same angular resolution of 15". In the high flux regime the PDF is proportional to the slice thickness.}
\label{thick_pdf}
\end{figure}

In this Section we investigate the effect of varying the thickness of the slices we cut through the simulation box on the shape of the surface brightness PDF. This allows us to test if the strongest emission, coming from haloes of different sizes, is entirely contained within a slice.
Results are shown in Fig. \ref{thick_pdf} for four different slice thicknesses (40, 20, 10 and 5 \hm\ Mpc). To facilitate the comparison, we have normalised the surface brightness PDF by the slice thickness $\Delta x$.

Within the potentially detectable regime the results are insensitive to the slice thickness. This occurs because the slices considered here are thick enough to fully contain the collapsed objects that give rise to the bulk of the UV emission, but at the same time they are thin enough that they avoid the superposition of multiple structures, which would strongly increase the flux in those pixels where more than one structure could be found along the line of sight.  At lower fluxes, the shape of the PDF depends on the flux in low density regions and on the number of low density structures such as filaments that can be found in each slice, which depends on the slice thickness. As a consequence, the width of the PDF at intermediate fluxes scales with the inverse of the slice thickness and the number of pixels at the lowest fluxes increases for decreasing slice thickness.

Throughout this work, we have assumed a slice thickness of 20 \hm\ Mpc. Fig. \ref{thick_pdf} demonstrate that this choice is well justified and allows us to estimate the flux from high density regions accurately. For fluxes higher than about 1 \phot\ the PDF is proportional to the slice thickness.

\bsp
\label{lastpage}

\end{document}